\documentclass[12pt,epsf]{article}
\usepackage{graphicx}
\usepackage{amssymb}

\begin{document}
%%%%%%%%%%%%%%%%%%%%%%%%%%%%%%%%%%%%%%%%%%%

\def\a{\alpha}
\def\b{\beta}
\def\c{\varepsilon}
\def\d{\delta}
\def\e{\epsilon}
\def\f{\phi}
\def\g{\gamma}
\def\h{\theta}
\def\k{\kappa}
\def\l{\lambda}
\def\m{\mu}
\def\n{\nu}
\def\p{\psi}
\def\q{\partial}
\def\r{\rho}
\def\s{\sigma}
\def\t{\tau}
\def\u{\upsilon}
\def\v{\varphi}
\def\w{\omega}
\def\x{\xi}
\def\y{\eta}
\def\z{\zeta}
\def\D{\Delta}
\def\G{\Gamma}
\def\H{\Theta}
\def\L{\Lambda}
\def\F{\Phi}
\def\P{\Psi}
\def\S{\Sigma}
\def\BR{{\rm Br}}
\def\o{\over}
\def\beq{\begin{eqnarray}}
\def\eeq{\end{eqnarray}}
\newcommand{\nn}{\nonumber \\}
\newcommand{\gsim}{ \mathop{}_{\textstyle \sim}^{\textstyle >} }
\newcommand{\lsim}{ \mathop{}_{\textstyle \sim}^{\textstyle <} }
\newcommand{\vev}[1]{ \left\langle {#1} \right\rangle }
\newcommand{\bra}[1]{ \langle {#1} | }
\newcommand{\ket}[1]{ | {#1} \rangle }
\newcommand{\EV}{ {\rm eV} }
\newcommand{\KEV}{ {\rm keV} }
\newcommand{\MEV}{ {\rm MeV} }
\newcommand{\GEV}{ {\rm GeV} }
\newcommand{\TEV}{ {\rm TeV} }
\def\diag{\mathop{\rm diag}\nolimits}
\def\Spin{\mathop{\rm Spin}}
\def\SO{\mathop{\rm SO}}
\def\O{\mathop{\rm O}}
\def\SU{\mathop{\rm SU}}
\def\U{\mathop{\rm U}}
\def\Sp{\mathop{\rm Sp}}
\def\SL{\mathop{\rm SL}}
\def\tr{\mathop{\rm tr}}

%added by FT
\newcommand{\bear}{\begin{array}}  
\newcommand {\eear}{\end{array}}
\newcommand{\la}{\left\langle}  
\newcommand{\ra}{\right\rangle}
\newcommand{\non}{\nonumber}  
\newcommand{\ds}{\displaystyle}
\newcommand{\red}{\textcolor{red}}
\def\ubl{U(1)$_{\rm B-L}$}
\def\REF#1{(\ref{#1})}
\def\lrf#1#2{ \left(\frac{#1}{#2}\right)}
\def\lrfp#1#2#3{ \left(\frac{#1}{#2} \right)^{#3}}
\def\OG#1{ {\cal O}(#1){\rm\,GeV}}

%%%%%%%%%%%%%%%%%%%%%%%%%%%%%%%
%%%    remove the following commands when finalizing
%%%%%%%%%%%%%%%%%%%%%%%%%%%%%%%
%\def\TODO#1{ {\bf ($\clubsuit$ #1 $\clubsuit$)} }
%%%%%%%%%%%%%%%%%%%%%%%%%%%%%%%
%%%%%%%%%%%%%%%%%%%%%%%%%%%%%%%

%%%%%%%%%%%%%%%%%%%%%%%%%%%%%%%%%%%%%%%%%%%%%%%%%%%%%%%%%%%%%%%%%%%%

\baselineskip 0.7cm

\begin{titlepage}

\begin{flushright}
UT-11-36\\
IPMU-11-0181
\end{flushright}

\vskip 1.35cm
\begin{center}
{\large \bf
A Solution to the $\mu/B\mu$ Problem in Gauge Mediation
with Hidden Gauge Symmetry
}
\vskip 1.2cm

{Koichi Hamaguchi$^{(a,b)}$, Kazunori Nakayama$^{(a,b)}$ and Norimi Yokozaki$^{(a)}$}

\vskip 0.4cm

{\it
$^a$Department of Physics, University of Tokyo, Bunkyo-ku, Tokyo 113-0033, Japan\\
$^b$Institute for the Physics and Mathematics of the Universe,
University of Tokyo, Kashiwa 277-8568, Japan\\
}

\vskip 1.5cm

\abstract{
We propose a solution to the $\mu/B\mu$ problem in gauge-mediated SUSY breaking, 
which does not suffer from the SUSY CP problem and is consistent with the solution to the strong CP problem. 
The model is based on $Z_3$-invariant NMSSM with additional vector-like matter
charged under a hidden gauge group as well as the standard model gauge groups.
The dynamical scale of the hidden gauge symmetry is set to be around 10 GeV.
 We show that this simple extension of the NMSSM resolves the domain wall problem and 
the $\mu/B\mu$ problem without a dangerous CP angle.
The relative sign among gaugino masses and the $\mu$ parameter can be preferable 
in terms of the muon anomalous magnetic moment. 
We also discuss cosmological issues of this model, especially the effects of
long-lived particles in the hidden gauge sector.
The hidden glueball may cause a late-time entropy production, which opens up a possibility that
the light gravitino can be a dominant component of dark matter while leptogenesis scenarios work successfully. There is a region where the electroweak symmetry is successfully broken with $\sim 10$ TeV stops, and the Higgs mass of $\sim 125$ GeV can be explained.
}
\end{center}
\end{titlepage}

\setcounter{page}{2}

%%%%%%%%%%%%%%%%%%%%%%%%%%%%%%%%%
\section{Introduction}
%%%%%%%%%%%%%%%%%%%%%%%%%%%%%%%%%

Gauge-mediated SUSY breaking (GMSB) is the one of the attractive mechanisms to transmit the SUSY breaking effects from the hidden sector to the visible sector~\cite{Giudice:1998bp}. 
In GMSB models there are no sources of CP violation and flavor violation in sfermion and gaugino sectors, since SUSY breaking effects are purely transmitted by gauge interactions. 
However, there are difficulties for generating viable $\mu$ term and $B\mu$ term. 
The first is the so-called $\mu/B\mu$ problem, which states that it is difficult to generate correct size of
both $\mu$ and $B$ parameters for the electroweak symmetry breaking in GMSB models.
The second difficulty is that there generally exists a CP violating phase among $\mu$ term, $B\mu$ term and gaugino masses.
Unless the relative CP phase is smaller than $\mathcal{O}(10^{-3})$,
it induces too large electric dipole moments of the electron and hadrons, 
which are excluded by the experiments~\cite{Nakamura:2010zzi}. 

There is an elegant solution to the $\mu/B\mu$ problem in the framework of Next-to-Minimal Supersymmetric Standard Model (NMSSM)~\cite{Ellwanger:2009dp}. 
By introducing additional vector-like matter charged under SM gauge groups to NMSSM, 
viable $\mu$ and $B\mu$ are generated without introducing an additional CP violating phase~\cite{Dine:1993yw,deGouvea:1997cx,Liu:2008pa,Morrissey:2008gm,Hamaguchi:2011nm}. 
It also eliminates the cosmological domain wall problem, which generally exists
due to the spontaneous breakdown of the $Z_3$ symmetry in NMSSM,
because the $Z_3$ symmetry is 
explicitly broken by the quantum anomaly induced by the vector-like matter~\cite{Preskill:1991kd}.
However, this solution is not compatible with the Peccei-Quinn (PQ) symmetry
for solving the strong CP-problem~\cite{Peccei:1977hh,Kim:1986ax}.
If this scenario is combined with the PQ mechanism,
a linear combination of the original $Z_3$ symmetry and the PQ symmetry 
leads to another non-anomalous $Z_3$ symmetry, and
the domain wall problem is restored~\cite{Preskill:1991kd,Hamaguchi:2011nm}.

In Ref.~\cite{Hamaguchi:2011nm} we constructed a model based on NMSSM with additional vector-like
matter charged under a {\it hidden} gauge group
and showed that such a model induces sizable $\mu/B\mu$-term, solves domain wall problem 
and is consistent with the PQ mechanism.
This model, however, was rather involved : it contained a hidden messenger sector,
which mediated the SUSY breaking effect into the hidden matter sector.

In this paper we propose a simpler extension of the NMSSM in GMSB.
We only introduce additional vector-like matter charged under both the SM and hidden gauge groups.
The messenger sector is as minimal as the the original GMSB model.
In this setup, we will show the $\mu/B\mu$ problem, the SUSY CP problem 
and the domain wall problem are all solved,
consistently with the PQ mechanism for solving the strong CP problem.
Moreover, the relative phase of the $\mu$-term and the gaugino masses can take 
a favored sign from the viewpoint of the SUSY explanation of the deviation of the 
muon $g-2$ (anomalous magnetic moment).

The novel difference from the previous model~\cite{Hamaguchi:2011nm} is that cosmology of the present model may be significantly modified from the standard one because of the existence of the long-lived particles
charged under both the SM and hidden gauge groups. 
%Since there exist long-lived particles in the hidden gauge sector, 
%they have non-trivial cosmological effects. 
Moreover, the hidden glueball may become a source
of a late-time entropy production, which might be welcome in GMSB where the upper bound on the 
reheating temperature after inflation is stringent.
The gravitino is a good dark matter candidate after the dilution by the decay of hidden glueball.

This paper is organized as follows.
In Sec.~\ref{sec:model} our model and its basic structure is described.
In Sec.~\ref{sec:mass} we describe phenomenology of the model, including
the mass spectrum, collider signatures, various constraints and implications on muon $g-2$.
In Sec.~\ref{sec:cos} cosmological aspects of this model is discussed in detail.
Sec.~\ref{sec:conc} is devoted to conclusions.

%%%%%%%%%%%%%%%%%%%%%%%%%
%
%
%
%
%%%%%%%%%%%%%%%%%%%%%%%%%%%%%%%%%
\section{NMSSM in GMSB with extra matter}   \label{sec:model}
%%%%%%%%%%%%%%%%%%%%%%%%%%%%%%%%%%%%%%
%

In this section we introduce a model of NMSSM in GMSB with extra matter,
which avoids the domain wall problem and is compatible with the PQ solution 
to the strong CP-problem.

\subsection{Model}

%%%%%%%%%%%%%%%%%%%%%%%%%%%%%%%%%%%%%%%%%%%%
%\begin{table}[t!]
%  \begin{center}
%    \begin{tabular}{  c | c | c | c | c | c | c | c | c | c | c | c  }
%     \hline 
%         ~          &  $S$  & $H_u$ & $H_d$ &${\bf 5^*_M}$ & ${\bf 10_M}$ & $\bar{D}'$ & $D'$ & ${\bar{L}'}$& $L'$ &$\Psi_{D,\bar{L}}$ & $\bar{\Psi}_{\bar{D},L}$\\
%       \hline \hline
%         $Z_3$  & $1$  & $1$ & $1$  & $1$   & $1$   & $0$ & $2$ & $0$ & $2$ & $0$ & $0$     \\ \hline 
%       R$_P$ &$+$ & $+$ & $+$  & $-$ & $-$ & $-$ & $-$ & $+$ & $+$ & $+$ & $+$   \\ \hline
%       $\SU(N_H)$ & ${\bf 1}$ & ${\bf 1}$ & ${\bf 1}$ & ${\bf 1}$ & ${\bf 1}$ & ${\bf N_H}^*$ & ${\bf N_H}$ & ${\bf N_H}$ & ${\bf N_H}^*$ & ${\bf 1}$ & ${\bf 1}$ 
%    \end{tabular}
%   \caption{
%     	Charge assignments on chiral superfields in the model under 
%	the $Z_3$, R-parity ($+$ : even, $-$ : odd) and the hidden gauge group SU$(N)_H$. 
%	${\bf 5^*_M}$ and ${\bf 10_M}$ are the MSSM matter fields.
%     }
%  \label{table:charge}
%  \end{center}
%\end{table}
%%%%%%%%%%%%%%%%%%%%%%%%%%%%%%%%%%%%%%%%%%%%%

%%%%%%%%%%%%%%%%%%%%%%%%%%%%%%%%%%%%%%%%%%%%%
\begin{table}[t!]
  \begin{center}
    \begin{tabular}{  c | c | c | c | c | c | c | c | c | c  }
%     \hline 
         ~          &  $S$  & $H_u$ & $H_d$ &${\bf 5^*_M}$ & ${\bf 10_M}$ & $\bar{D}'({L'})$ & ${D}'({\bar{L}'})$ &$\Psi_{D,\bar{L}}$ & $\bar{\Psi}_{\bar{D},L}$\\
       \hline \hline
         $Z_3$  & $1$  & $1$ & $1$  & $1$   & $1$   & $1$ & $1$ & $0$ & $0$      \\ \hline 
       R$_P$ &$+$ & $+$ & $+$  & $-$ & $-$ & $-$ & $-$ & $-$ & $-$   \\ \hline
       $\SU(N_H)$ & ${\bf 1}$ & ${\bf 1}$ & ${\bf 1}$ & ${\bf 1}$ & ${\bf 1}$ & ${\bf N_H}^*$ & ${\bf N_H}$ & ${\bf 1}$ & ${\bf 1}$ 
    \end{tabular}
    \caption{ 		
     	Charge assignments on chiral superfields fields in the model under 
	the $Z_3$, R-parity ($+$ : even, $-$ : odd) and SU$(N)_H$. 
	${\bf 5^*_M}$ and ${\bf 10_M}$ are the MSSM matter fields.
     }
  \label{table:charge}
  \end{center}
\end{table}
%%%%%%%%%%%%%%%%%%%%%%%%%%%%%%%%%%%%%%%%%%%%%

We consider the following $Z_3$ invariant superpotential:
\begin{eqnarray}
W =  W_{\rm NMSSM} + W_{\rm GMSB}  + W_{\rm extra}
\label{eq:sp1}
\end{eqnarray}
where
\begin{eqnarray}
W_{\rm NMSSM} &=& \lambda S H_u H_d + \frac{\kappa}{3} S^3 + W_{\rm MSSM\!-\!Yukawa}\,,
\\
W_{\rm GMSB} &=& X \Psi \bar{\Psi} + W_{\rm hid}(X)\,,
\\
W_{\rm extra} &=& k_{D'} S D' \bar{D}' + k_{L'} S L' \bar{L}'\,.
\end{eqnarray}
The charge assignments of chiral superfields are shown in Table \ref{table:charge}.
$W_{\rm NMSSM}$ is the superpotential of the NMSSM, where singlet Higgs, up-type Higgs and down-type Higgs are denoted by $S$, $H_u$ and $H_d$, respectively. The effective $\mu$-term is 
induced by the vacuum expectation value (VEV) of the singlet field, $\mu_{\rm eff}=\lambda \vev{S}$.
$W_{\rm GMSB}$ is responsible for the gauge mediation of the SUSY breaking,
where messenger superfields are denoted by $\Psi$ and $\bar{\Psi}$, 
which are ${\bf 5}$ and ${\bf 5^*}$ representation of $\SU(5)$ grand unified theory gauge group. 
The SUSY breaking field $X$ is assumed to develop a VEV as $X=M_{\rm mess}+\theta^2 F_{X}$ due to the dynamics induced by $W_{\rm hid}(X)$ (and K\"ahler potential).

Finally, $\bar{D}'$ and ${L}'$ (${D}'$ and $\bar{L}'$) are extra vector-like matter, which are charged under 
the SM gauge group as well as an $\SU(N_H)$ hidden gauge group. 
They naturally induce the VEV of the $S$ field through the effects of renormalization group evolution on the mass of $S$, 
and also play an important role to break the $Z_3$ symmetry via anomaly.
As will be discussed, the hidden gauge group $\SU(N_H)$ is necessary to break the $Z_3$ symmetry 
in the presence of the PQ mechanism.
Note that $N_H$ should be either 2 or 3, since the perturbative gauge coupling unification fails for larger $N_H$.  
As we will see, it turns out that the dynamical scale of the hidden gauge group, 
$\Lambda_H$, should be about 10 GeV.
The mass spectrum of the model is discussed in Sec.~\ref{sec:mass}.

%%%%%%%%%%%%%%%%%%%%%%%%%%%%%%%%%
\subsection{Domain wall and strong CP problem}
\label{sec:DWandPQ}
%%%%%%%%%%%%%%%%%%%%%%%%%%%%%%%%%

It is known that the $Z_3$ invariant NMSSM suffers from the domain wall problem
since the $Z_3$ symmetry is spontaneously broken after the electroweak phase transition.
One may introduce a small explicit $Z_3$ breaking term in the superpotential,
but this in general leads to a tadpole term which tends to 
destabilze the gauge hierarchy~\cite{Abel:1995wk,Abel:1996cr}.\footnote{
See e.g., ~\cite{Abel:1996cr, Panagiotakopoulos:1998yw, Panagiotakopoulos:1999ah, Lee:2011dya} for models without the tadpole problem.
}

In the present model, however, the $Z_3$ is anomalous at the quantum level due to the additional vector-like matter.
By integrating out $D'$, $\bar{D}'$, $L'$ and $\bar{L}'$, a $Z_3$ breaking term is induced as
\begin{eqnarray}
\mathcal{L} = \frac{g_H^2}{64\pi^2} \left(\theta_H + N\frac{a_S}{v_S}\right) \epsilon^{\mu\nu\rho\sigma}{G'}^a_{\mu\nu} {G'}^a_{\rho\sigma},
\end{eqnarray}
where $G_{\mu\nu}'$ is the hidden gluon field strength and $\theta_H$ is the sum of the strong phase of the hidden gauge group $\SU(N_H)$ and phases induced by the mass of extra-matter,
which is in general expected to be $\mathcal O(1)$, and $N$ 
counts the number of matter fields with 
fundamental and anti-fundamental representation of $\SU(N_H)$ ($N=5$ in our model). 
We denote a singlet CP-odd Higgs by $a_S$, and $v_S=\left<S\right>$ is the VEV of $S$.
Apparently, this term violates the $Z_3$ symmetry while the rest of the Lagrangian is invariant. 
After the hidden QCD phase transition, 
there arises a potential for $a_S$ which explicitly breaks $Z_3$ symmetry~\cite{Shifman:1979if}
\begin{eqnarray}
	V_{\rm Z_3 \hspace{-8pt}/} \sim \Lambda_{H}^4 f\left( \theta_H + N a_S/v_S \right), \label{eq:Z3b}
\end{eqnarray}
where $f(x)$ is a periodic function which satisfies $f(x+2\pi)=f(x)$ and  
$\Lambda_H$ is the strong scale of the hidden gauge interaction~\cite{Witten:1979vv}.
Therefore, by comparing it with the tree level $Z_3$-invariant scalar potential, we see that
the degeneracy of the three distinct vacua are broken if an integer $N$ and $3$ are relatively prime. 
The bias among the potential energies of the original three vacua are $\Delta V \sim \Lambda_H^4$.
As explicitly shown in Ref.~\cite{Hamaguchi:2011nm},
$\Lambda_H \gsim \mathcal{O}(1)$~MeV is
sufficient to make domain walls unstable so that they decay well before dominating the Universe.
Note that the potential (\ref{eq:Z3b}) slightly shifts the position of the minimum of $a_S$ 
from zero, which induces a small CP violation. We will discuss it in Sec.~\ref{sec:smallCP}.

Here let us see that our model is compatible 
with the PQ solution to the strong CP problem~\cite{Peccei:1977hh,Kim:1986ax}.
Since the MSSM sector does not exhibit a PQ symmetry, 
we introduce a new sector in the superpotential,
\begin{equation}
	W =  k \Phi_{\rm PQ} Q_{\rm PQ}\bar Q_{\rm PQ},
\end{equation}
where $\Phi_{\rm PQ}$ is a gauge singlet and $Q_{\rm PQ}$ ($\bar Q_{\rm PQ}$) 
are (anti-)fundamental representations of $\SU(5)$.\footnote{
	We introduce only one set of such matter pairs.
	If more matter is introduced, the gauge coupling may blow up below the GUT scale.
	Note also that color anomaly number is equal to one for the model with only one pair,
	and hence there is no axionic domain wall problem 
	even if the PQ symmetry is restored after inflation~\cite{Vilenkin:1982ks}.
}
The global U(1)$_{\rm PQ}$ charges are assigned as $\Phi_{\rm PQ} (+2), Q_{\rm PQ}(-1), \bar Q_{\rm PQ}(-1)$.
After $\Phi_{\rm PQ}$ develops a VEV of $f_a \sim \mathcal O(10^{10}-10^{12})$\,GeV,
the U(1)$_{\rm PQ}$ is spontaneously broken and then an almost massless mode
corresponding to the angular component of $\Phi_{\rm PQ}$, axion ($a_{\rm PQ}$), appears.
In this paper we do not specify the mechanism of the stabilization of the PQ scalar
along the flat direction associated with the U(1)$_{\rm PQ}$ symmetry.
Although it is irrelevant for the following discussions, we here show one example.
A simple way is to introduce a superpotential of the form $W = \Phi_{\rm PQ}^n \bar\Phi_{\rm PQ}/M^{n-2}$
with a cutoff scale $M$ where $\bar \Phi_{\rm PQ}$ has a PQ charge $-2n$.
Then the PQ scalar is stabilized at $\langle \Phi_{\rm PQ}\rangle \sim (m_{\rm PQ} M^{n-2})^{1/(n-1)}$
and $\langle \bar\Phi_{\rm PQ}\rangle = 0$,
where $m_{\rm PQ}$ is the SUSY breaking mass for the PQ scalar.
By choosing $n$ appropriately, we can obtain a desired PQ scale $f_a = \langle \Phi_{\rm PQ}\rangle$.

The U(1)$_{\rm PQ}$ is anomalous under QCD.
Since the $Z_3$ is also anomalous under QCD, the instanton effects induce 
the following potential,
\begin{equation}
	V_{\rm PQ} \sim \Lambda_{\rm QCD}^4 
	 \left[ 1-\cos\left(a_{\rm PQ}/f_a+N_H a_S/v_S + \theta_0 \right) \right],   \label{VQCD}
\end{equation}
where $\theta_0$ is the bare strong CP angle.
Independently of the value of $a_S$, the axion always
dynamically cancels the CP angle and resolves the strong CP problem.
Notice that, without a $Z_3$-breaking potential induced by hidden gauge group (\ref{eq:Z3b}), 
we could always choose $a_{\rm PQ}$ so that $V_{\rm PQ} = 0$ for each $Z_3$ vacuum,
i.e., there would remain an unbroken $Z_3$ symmetry.
This is why we need a strong hidden gauge symmetry to make the $Z_3$ anomalous.

%%%%%%%%%%%%%%%%%%%%%%%%%%%%%%%%%
\section{Mass spectrum and phenomenology}   \label{sec:mass}
%%%%%%%%%%%%%%%%%%%%%%%%%%%%%%%%%

Let us now discuss the mass spectrum and  phenomenology of the model.\footnote{
The analyses in this section are out of date in light of the recent discovery of the 125GeV Higgs boson
after the submission of this paper. See ``Note Added'' for updated analyses. 
}
The model is parameterized by the following parameters:
\begin{equation}
M_{\rm mess},\; 
\Lambda=\frac{F_{X}}{M_{\rm mess}},\;
\lambda,\;
\kappa,\;
k_{D'},\;k_{L'},\;
N_H,\; g_H .
\end{equation}
In our analysis, we fix $\kappa$ and $k_{D'}$ in terms of $\tan\beta$ and the Higgs VEV, similar to the case of MSSM
where $\mu$ and $B\mu$ are fixed. 
As described below, the mass squared of the gauge singlet, $m_S^2$, 
which should be negative for successful EWSB,  is controlled by the parameter $k_{D'}$.
The parameter $k_{L'}$ does not affect the EWSB as significantly as $k_{D'}$ does. Therefore in numerical calculation, we fix it to be a small value, 
since otherwise there appears an unbounded vacuum, as described in Sec.~\ref{sec:vacuum}.
$N_H$ should be 2 or 3 in order to keep the perturbative coupling unification.
As will be discussed, viable phenomenology and cosmology require
$\Lambda_H\sim 10$~GeV, which corresponds to $g_H(1 {\rm TeV}) \simeq 1.2\,  (1.4)$ for $N_H=3\,(2)$.
Thus, there are essentially 4 parameters left, $M_{\rm mess}$, $\Lambda$, $\lambda$, and $\tan\beta$.

Below the messenger scale, the following soft terms are induced,
\begin{eqnarray}
-\mathcal{L}_{\rm soft} &=& m_S^2 |S|^2 + m_{H_u}^2 |H_u|^2 + m_{H_d}^2 |H_d|^2 \nn
&&  +m_{D'}^2 |D'|^2 +m_{\bar{D}'}^2 |\bar{D}'|^2 +  m_{L'}^2 |L'|^2 +m_{\bar{L}'}^2 |\bar{L}'|^2 \nn
&& + (A_\lambda \lambda S H_u H_d + A_\kappa \kappa S^3/3 + h.c. ) \nn
&& + (A_{D'} k_{D'} S D' \bar{D'} + A_{L'} k_{L'} S L' \bar{L}' + h.c.) .
\end{eqnarray}
where the soft terms of the squarks, sleptons, and gauginos are omitted for simplicity.
The soft masses for the extra matter, $m_{D'}^2$, $m_{\bar{D}'}^2$, $m_{L'}^2$, $m_{\bar{L}'}^2$ are generated in
a similar way to the MSSM squarks and sleptons via GMSB mechanism,
since they are charged under the SM gauge groups.
In the presence of this extra matter, the beta-function for $m_S^2$ receives additional contributions, 
\begin{eqnarray}
(8\pi^2)\frac{d m_S^2}{d \ln Q} &\ni&  3 N_H k_{D'}^2 (m_{D'}^2 + m_{\bar{D}'}^2 + m_S^2 + |A_{D'}|^2) \nn
&& + 2 N_H k_{L'}^2 (m_{L'}^2 + m_{\bar{L}'}^2 + m_S^2 + |A_{L'}|^2)
\end{eqnarray}
where $Q$ is the renormalization scale. This induces a negative $m_S^2$ and the VEV of $S$ field accordingly, 
thereby solving the $\mu$/$B\mu$ problem.
Since $m_{D'}^2$ and $m_{\bar{D}'}^2$ are larger than $m_{L'}^2$ and $m_{\bar{L}'}^2$ in GMSB, the $m_S^2$ and $\langle S\rangle$ are mainly controlled by the parameter $k_{D'}$.

In the numerical calculations, we have used the NMSSM Tools~\cite{Ellwanger:2008py},
which is modified to include the effects of additional vector-like fields,\footnote{
$k_{D'}$ is determined iteratively so that the conditions for the EWSB are satisfied with predicted soft mass parameters.
} 
and to include two-loop effects of additional hidden gauge group (see Appendix A). 
The mass spectrum, the LEP constraint and the branching ratio of the inclusive B meson decay are calculated in NMSSM Tools.

Now let us see various constraints and implications of the model.

%%%%%%%%%%%%%%%%%%%%%%%%%%%%%%%%%
\subsection{NMSSM mass spectrum}
%%%%%%%%%%%%%%%%%%%%%%%%%%%%%%%%%

%%%%%%%%%%%%%%%% table %%%%%%%%%%%%%%%%%%%%%%
\begin{table}[t]
  \begin{flushleft}
    \begin{tabular}{ | c | c | c | c | c | c | c | c | c | c |}
      \hline 
         ~          &  $\Lambda$  & $M_{\rm mess}$ & $g_H(m_{\tilde{q}})$ & $k_{D'}$ & $N_H$ & $\lambda$ & $\kappa$ &$\tan\beta$ & $\mu_{\rm eff}$    \\     \hline \hline
         P1  & $1.2 \times 10^5$ & $10^9$ & $1.48$  & $ 3.40\times 10^{-2}$   & $2$   & $0.006$ & $-2.28\times10^{-4}$ & $42$ & $804$ \\ \hline 
         P2  & $1.0\times 10^5$  & $10^{9}$ & $1.20$    & $2.99\times 10^{-2}$   & $3$   & $0.005$ & $-2.01\times10^{-4}$ & $37$ & $831$ \\ \hline 
         P3  & $1.4 \times 10^5$ & $10^8$ & $1.13$  & $ 2.22\times 10^{-2}$   & $3$   & $0.005$ & $-1.56\times10^{-4}$ & $45$ & $1006$ \\ \hline 
    \end{tabular}
    \begin{tabular}{ | c | c | c | c | c | c | c | c | c | c | c |c|}
      \hline 
         ~          &  $m_{h_1}$  & $m_{h_2}$  & $m_{a_1}$ &$m_{a_2}$ & $m_{\chi_1^0}$ & $m_{\chi_1^+}$ & $\tilde{\tau}_{1}$  & $\tilde{t}_1$ & $\tilde{q}_{1,2}$ &  $\tilde{g}$ & $\Delta a_\mu$\\
       \hline \hline
         P1  & $58.6$  & $116.8$ & $7.9$  & $681.6$   & $61.3$  & $313.2$ & $193.8$ & $1210.5$ & $1480$ & $941.1$ & $2.14\times 10^{-9}$\\ \hline 
        P2  & $65.0$  & $116.6$ & $9.1$  & $744.6$   & $67.0$   & $260.6$ & $176.8$ & $1223.9$ & $1490$ & $793.3$ & $2.27\times 10^{-9}$ \\ \hline
        P3  & $60.3$  & $118.3$ & $6.6$  & $772.8$   & $63.0$  & $367.9$ & $216.4$ & $1641.1$ & $1930$ & $1084.1$ & $1.69\times 10^{-9}$\\ \hline 
%     
%       SU(2)$_L$ &${\bf 1}$& ${\bf 2}$ & ${\bf 2}$ & ${\bf 1}$ & ${\bf 1}$ &${\bf 2}$& ${\bf 2}$& ${\bf 1+2}$ & ${\bf 1+2}$ & ${\bf 1}$   \\ \hline
    \end{tabular}
        \begin{tabular}{ | c | c | c | c | c|c|c|}
      \hline 
         ~          &  $m_{\psi_{L'}}$  & $m_{\psi_{D'}}$  & $m_{\tilde{L}'_1}$  & $m_{\tilde{L}'_2}$& $m_{\tilde{D}'_1}$  & $m_{\tilde{D}'_2}$\\
       \hline \hline
         P1  & $267.9$  & $4559.9$ & 571.4 & 617.6 & 4411.0 & 5107.2\\ \hline 
        P2  & $332.4$  & $4966.9$  & 561.8 & 619.9 & 4794.2 & 5515.0 \\ \hline 
         P3  & $402.4$  & $4471.9$ & 713.7 & 771.8 & 4460.1 & 5198.0\\ \hline 
%     
%       SU(2)$_L$ &${\bf 1}$& ${\bf 2}$ & ${\bf 2}$ & ${\bf 1}$ & ${\bf 1}$ &${\bf 2}$& ${\bf 2}$& ${\bf 1+2}$ & ${\bf 1+2}$ & ${\bf 1}$   \\ \hline
    \end{tabular}
    \caption{The mass spectra of some model points are shown. The input parameters are $\Lambda$, $M_{\rm mess}$, $\lambda$, $\tan\beta$, $g_H$ and $N_H$. The parameters, $k_{D'}$, $\kappa$ and $\mu_{\rm eff}$ are determined by iteration. $k_{L'}$ is taken as $k_{L'}=0.002$ for all numerical calculations.
    The fermion masses of $L'$ and $D'$ are denoted by $m_{\psi_{L'}}$ and $m_{\psi_{D'}}$ respectively.
    All masses are written in units of GeV.
     }
     \label{tab:mass}
  \end{flushleft}
\end{table}

Typical viable parameter regions are shown in Fig.~1--4 and the mass spectrum of some model points are shown in 
Table~\ref{tab:mass}. 
Various constraints and phenomenological implications of the model will be discussed in more detail in the following subsections.
In the viable region we are interested in, the SUSY breaking scale is $F_X\gsim 10^{12}\,  {\rm GeV}^2$
and hence the gravitino is heavier than $\mathcal O(1)\,{\rm keV}$. 
The lightest SUSY particle within the NMSSM sector is the lightest neutralino, which is stable inside the detector.
In these cases, SUSY searches limit the squark masses, $m_{\tilde{q}}$ as $m_{\tilde{q}} \gtrsim 1100$ GeV for the gluino mass $m_{\tilde{g}} \simeq \, 900 {\rm GeV}$, and $m_{\tilde{q}} \gtrsim 1200$ GeV for $m_{\tilde{g}} \simeq  \, 800 {\rm GeV}$~\cite{Aad:2011ib, Chatrchyan:2011zy}.
Figs.~{\ref{fig:mgluino}} and {\ref{fig:msquark}} show contours of the gluino mass and squark mass, respectively. 
There are viable regions where the constraints from the SUSY searches are avoided while the deviation of the muon $g-2$
from the SM prediction is within 1$\sigma$ level (see also P1, P2 and P3 in Table~\ref{tab:mass}). 
Note that for $N_H=3$, the one-loop renormalization group coefficient of the gluino mass vanishes and the two-loop contribution is important for deriving the mass spectrum. (See Appendix A).

In typical viable points, our model predicts that the lightest CP-even Higgs, $h_1$ is singlet like and is as light as $\mathcal O(10)$ GeV.\footnote{
	The solution with $\mu_{\rm eff}<0$ predicts heavier Higgs masses (see Refs.~\cite{Morrissey:2008gm,Hamaguchi:2011nm}). 
} 
Therefore it is excluded by the LEP results if its mixing with the SM-like Higgs is too large. 
Since larger $\lambda$ leads to larger mixings with $H_u$ and $H_d$, there is an upper bound on $\lambda$
to avoid the LEP constraints on the mass of $h_1$ and the coupling, $Z-Z-h_1$~\cite{Barate:2003sz}.
In Fig.~\ref{fig:vac_lep}, we show the constraint on $\lambda$ on $\lambda$-$\Lambda$ plane. 
In order to avoid the LEP constraint, $\lambda$ should be as small as $\mathcal O(10^{-3})$. 
On the other hand, the second lightest CP-even Higgs is SM-like, 
and its mass about $117$\,GeV by including top/stop radiative corrections.\footnote{
	We have checked that FeynHiggs~\cite{Hahn:2010te} also predicts the SM-like Higgs boson mass 
	of $\simeq116.5$ GeV, with the parameters in P1 and P2.
}

In most of the viable regions, the lightest CP-odd Higgs is singlet like and is as light as a few GeV. 
The lightness originates from the smallness of $\kappa\, (\sim 10^{-4})$ and $A_\kappa$, which corresponds to an approximate PQ symmetry.\footnote{
	This is different from the PQ symmetry discussed in Sec.~\ref{sec:DWandPQ} for the solution to the strong CP problem.
}  Smallness of $\kappa$ is guaranteed from the requirement that the correct electroweak symmetry breaking is obtained when $\lambda \ll 1$ (i.e. $\left< S\right> \gg 1 $TeV) and the soft mass parameters, $|m_{H_d}^2|$ and $|m_S^2|$, are not much larger than $10^6\, {\rm GeV}^2$. Such a very light CP-odd Higgs can affect the prediction of ${\rm Br}(B \to X_s \mu^+ \mu^-)$ significantly, as described Sec.~\ref{B phys}.

%%%%%%%%%%%%%%%
\begin{figure}[tbp]
\begin{center}
\vspace{-0.5cm}
\includegraphics[scale=0.9]{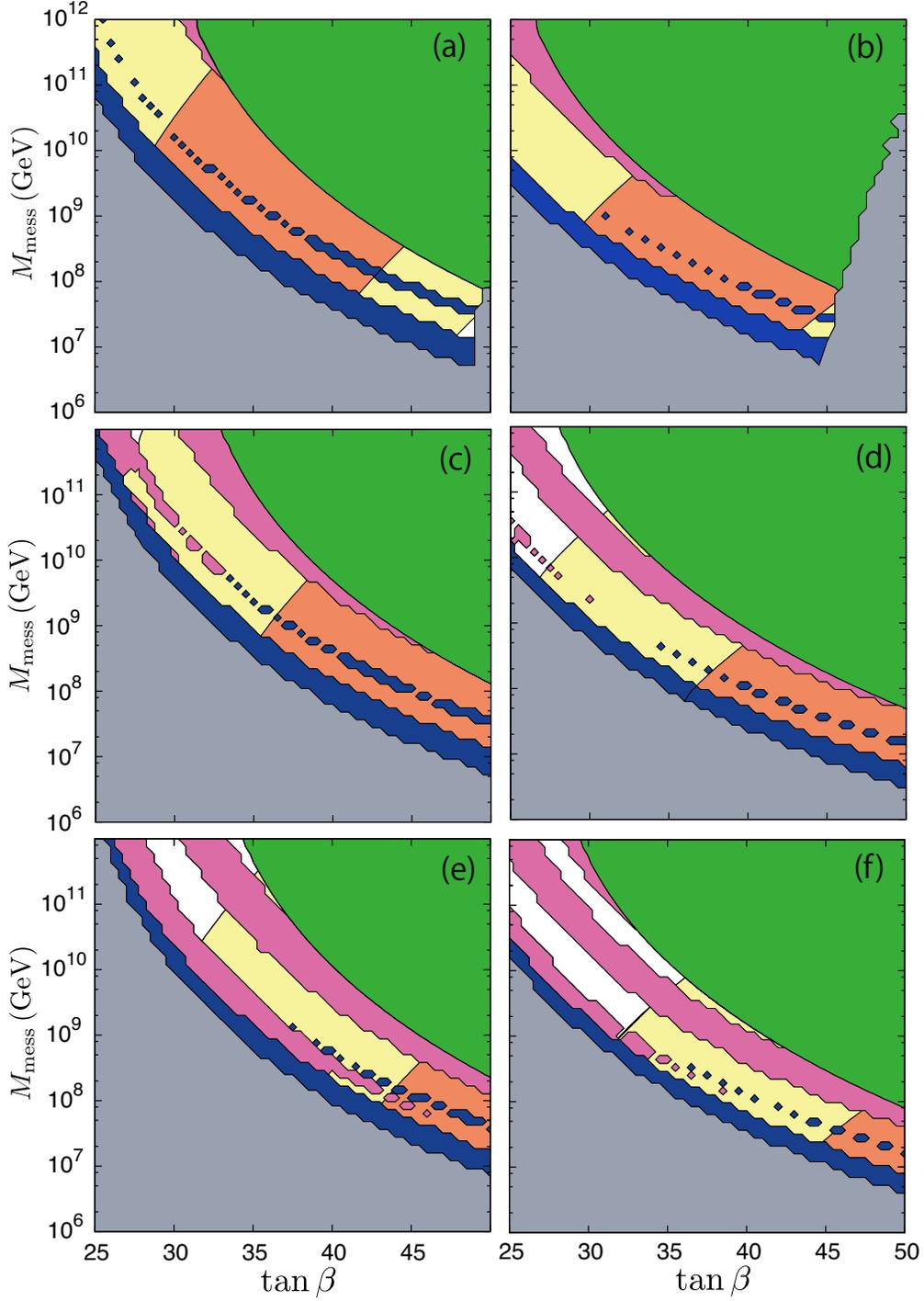}
\caption{ The contours of g-2 and allowed regions of the parameters. $\Lambda$=100 TeV, 120TeV and 140TeV for (a)(b), (c)(d) and (e)(f), respectively. In panels (a)(c)(e), $\lambda=0.006$ and $N_H=2$, and in  panels (b)(d)(f),  $\lambda=0.005$ and $N_H=3$. 
In orange (yellow) regions, muon g-2 is explained at 1$\sigma$ (2$\sigma$) level.
Pink regions are excluded by the LEP constraint. Blue regions are excluded by $B \to X_s \mu^+ \mu^-$. Green regions are excluded by unbounded vacuum. Gray regions are not consistent with the experimental data of the muon g-2, since there is no solution with $\mu_{\rm eff}>0$.
The contours are drawn with $g_H(m_{\tilde{q}})=1.2$, where $m_{\tilde{q}}$ is a squark mass.
}
\label{fig:cont}
\end{center}
\end{figure}
%%%%%%%%%%%%%%%

%%%%%%%%%%%%%%%
\begin{figure}[tbp]
\begin{center}
 \includegraphics[scale=0.98]{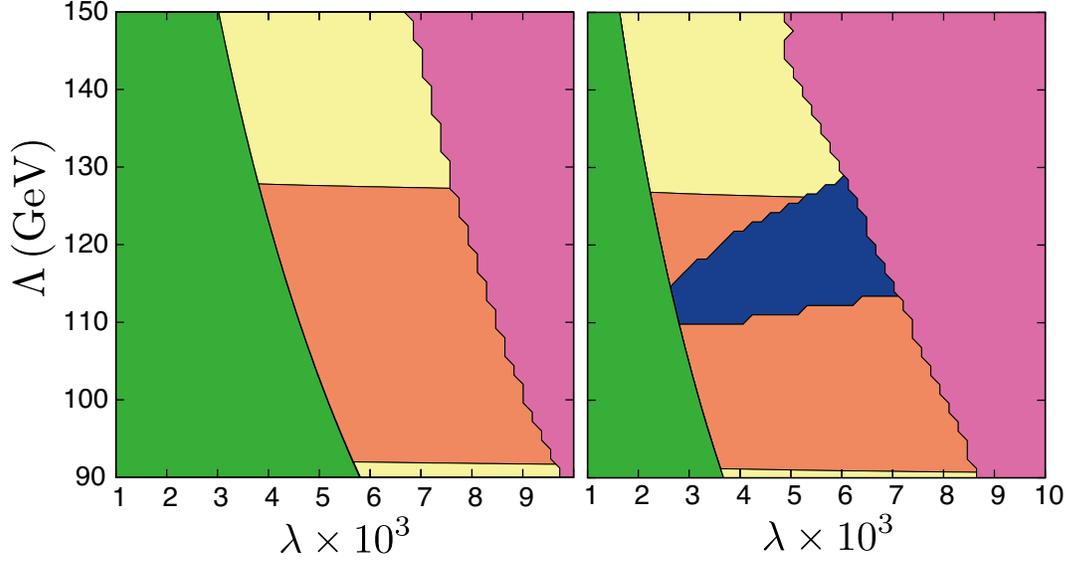}
\caption{ The allowed region of $\lambda$ is shown. In the left panel, we take $N_H=2$, $\tan\beta=40$ and $M_{\rm mess}=10^9 {\rm GeV}$ and in the right panel, $N_H=3$, $\tan\beta=40$ and $M_{\rm mess}=10^8 {\rm GeV}$. $g_H(m_{\tilde{q}})$ is taken as $g_H(m_{\tilde{q}})=1.2$. The meanings of the colored regions are same as those in Fig. 1.
%The pink regions are excluded  due to the LEP constraint while blue regions are excluded by $B \to X_s \mu^+ \mu^-$.
}
\label{fig:vac_lep}
\end{center}
\end{figure}
%%%%%%%%%%%%%%%

%%%%%%%%%%%%%%%
\begin{figure}[tbp]
\begin{center}
 \includegraphics[scale=1.0]{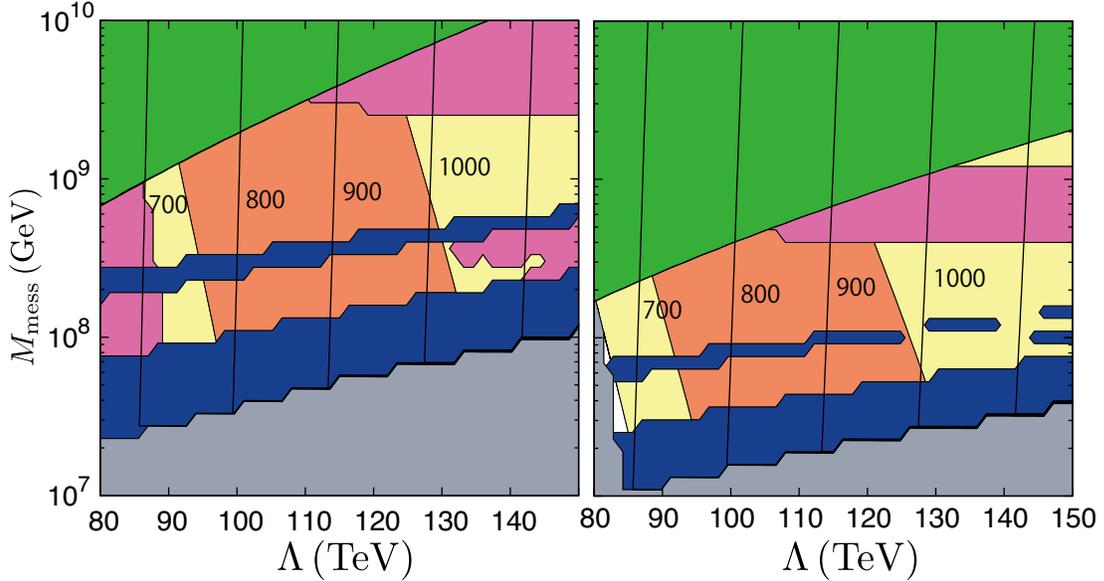}
\caption{ The contours of gluino mass are shown. The numbers are expressed in the unit of GeV. In the left panel, $N_H=2$, $\lambda=0.006$ and $\tan\beta=40$ and in the right panel, $N_H=3$, $\lambda=0.005$ and $\tan\beta=40$. $g_H$ is taken as
$g_H(m_{\tilde{q}})=1.2$.  The meanings of the colored regions are same as those in Fig. 1.}
\label{fig:mgluino}
\end{center}
\end{figure}
%%%%%%%%%%%%%%%

%%%%%%%%%%%%%%%
\begin{figure}[tbp]
\begin{center}
 \includegraphics[scale=1.0]{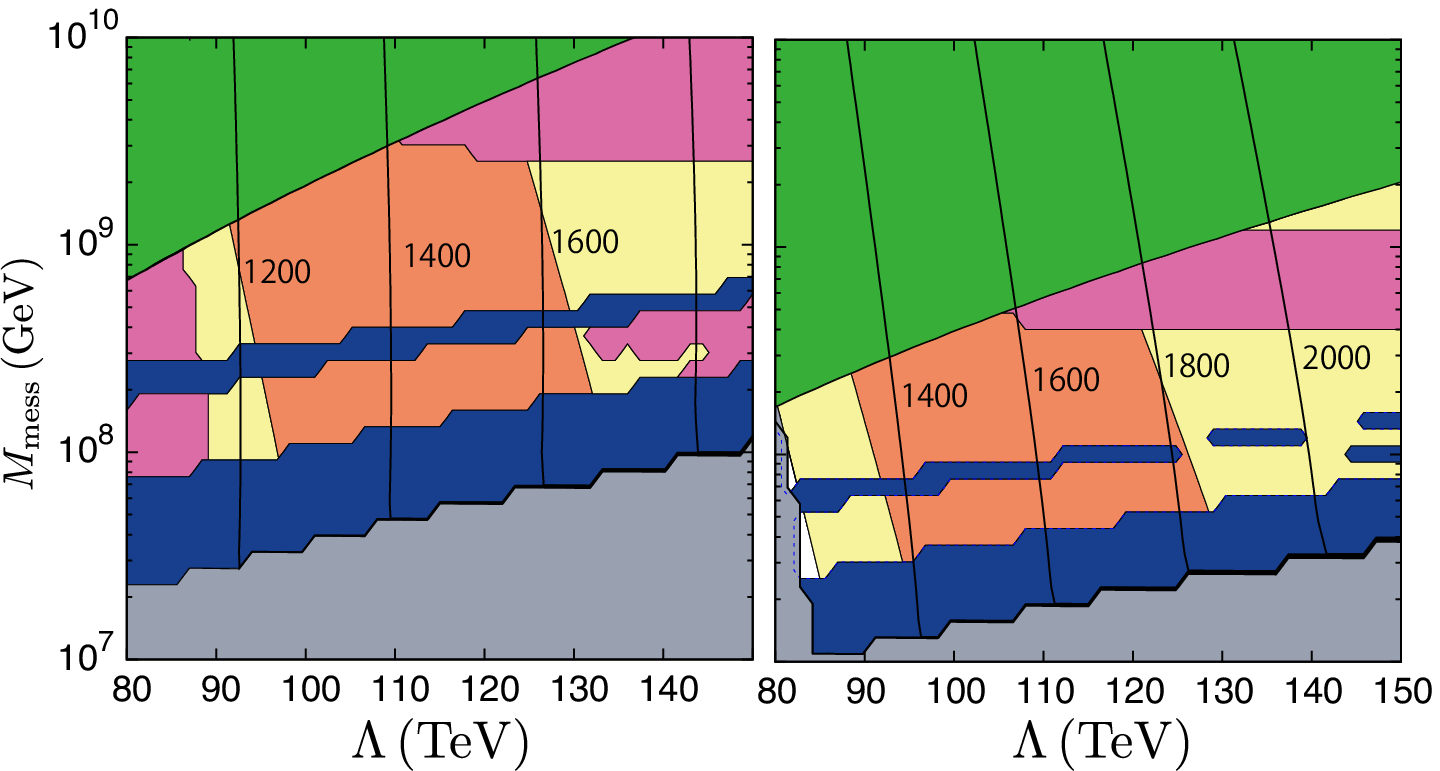}
\caption{ The contours of squark mass are shown. The numbers are expressed in the unit of GeV. The parameters and colored regions are same as in Fig.~{\ref{fig:mgluino}}}
\label{fig:msquark}
\end{center}
\end{figure}
%%%%%%%%%%%%%%%

%%%%%%%%%%%%%%%%%%%%%%%%%%%%%%%%%
\subsection{Anomalous Magnetic Moment of the Muon}  \label{sec:g-2}
%%%%%%%%%%%%%%%%%%%%%%%%%%%%%%%%%

The experimental value of the muon $g-2$ deviates from the SM prediction with about 3$\sigma$ level~\cite{Hagiwara:2011af, Davier:2010nc}. In Ref.~\cite{Hagiwara:2011af} it is calculated as
\begin{eqnarray}
(a_\mu)_{\rm EXP} - (a_\mu)_{\rm SM} = (26.1 \pm 8.0) \cdot 10^{-10}
\end{eqnarray} 
where the subscripts EXP and SM refer to the experimental value and the SM prediction.
Interestingly, the deviation can be naturally explained by SUSY contributions when $\tan\beta$ is $\mathcal O(10)$~\cite{Chattopadhyay:1995ae}. 
The SUSY contribution to the muon $g-2$ is approximately given by~\cite{Moroi:1995yh}\footnote{
	This approximation is valid only when the diagram in which $\tilde{\nu}_\mu$ and $\tilde{\chi}^-$
	dominantly contribute to the loop among other SUSY contributions.
}
\begin{eqnarray}
\Delta a_{\mu} \sim \frac{g_2^2}{32\pi^2} \frac{m_\mu^2}{m_{\rm soft}^2} \tan\beta \frac{M_2 \mu}{|M_2 \mu|}.
\end{eqnarray}
It is enhanced for large $\tan\beta$ and suppressed for large $m_{\rm soft}^2$. 
In order to explain the muon g-2 anomaly, the sparticles (sleptons) should be sufficiently light.
Note that the relative sign between $M_2$ and $\mu$ is also important. 
For $M_2 \mu < 0$, SUSY contributions tend to make $a_\mu$ more discrepant from the SM prediction.
In Fig.~1-4, contours of $g-2$ are shown. In the orange (yellow) regions, 
the muon $g-2$ agrees with the experimental result within 1$\sigma$ (2$\sigma$) level.
The viable region consistent with the muon $g-2$ corresponds to $F_X\simeq10^{12}-10^{16}\, {\rm GeV}^2$. 
Therefore, the gravitino mass is $m_{3/2} \gsim F_X/M_P \simeq \mathcal{O}(0.001-10)\, {\rm MeV}$. Coincidentally, such a light gravitino is also preferable from the viewpoint of cosmology as described in Sec.~\ref{sec:cos}. 

Since SUSY searches excluded the light sparticle mass regions in GMSB, 
a large $\tan\beta$ is necessary in order to explain the muon $g-2$.
For large $\tan\beta$, only viable region with successful electroweak symmetry breaking is 
$|\kappa/\lambda| \ll1$ and $\lambda \ll 1$ as long as soft masses are induced by GMSB effects~\cite{Morrissey:2008gm}. 
This is the so-called decoupling limit of the NMSSM. 
However, very small values of $\lambda$ are also constrained from the vacuum structure, as we will see below.

%%%%%%%%%%%%%%%%%%%%%%%%%%%%%%%%%
\subsection{Vacuum structure}
\label{sec:vacuum}
%%%%%%%%%%%%%%%%%%%%%%%%%%%%%%%%%

In the NMSSM, the vacuum structure is complex compared to that of MSSM, due to the presence of 
another singlet Higgs; there exist unwanted local (global) minima in general (c.f.~\cite{Kanehata:2011ei}). Therefore the depth of these minima should be compared to that of the wanted minimum so that the successful electroweak symmetry breaking is realized. 

The relative depth of the wanted minimum, which we call ``true'' minimum hereafter, 
measured from the origin $S=H_u=H_d=0$ is roughly given by,
\begin{eqnarray}
	V_{\rm true} \simeq -(m_S^2)^2/(4\kappa^2) \sim -10^{14}{\rm GeV}^4,
\end{eqnarray}
where we neglect small $A$-term contributions. 
Since we consider the decoupling limit where $v_S$ is as large as $\mathcal O(100)$ TeV, 
the VEVs of $H_u$ and $H_d$ are approximately neglected in the following analysis.
Now let us see other local or global minima existing in the present model.

%%%%%%%%%%%%%%%%%%%%%%%%%%%%%%%%%
\subsubsection{Avoiding the color breaking minimum}\label{CCB}
%%%%%%%%%%%%%%%%%%%%%%%%%%%%%%%%%

First note that there may be additional constraints on the vacuum structure once
the extra matter is included~\cite{Hamaguchi:2011nm}.
The mos stringent constraint comes from requiring the absence of unbounded direction. 
The unbounded direction exists when the soft masses of $D'$ and $\bar{D}'$ and/or $k_{D'}$ are small.
Along the direction of $\left<S\right>=0$ , $|\left<H_u^0\right>|=|\left<H_d^0\right>|=v_{H}$ and $|\left<D'\right>|=|\left<\bar{D}'\right>|=v_{D}$, the scalar potential reads
\begin{eqnarray}
	V= \Bigl(-|\lambda| v_H^2 + |k_{D'}| v_{D'}^2 \Bigr)^2 + (m_{H_u}^2+m_{H_d}^2)v_H^2 + (m_{D'}^2 + m_{\bar{D}'})v_D^2.
\end{eqnarray}
The first term vanishes for $v_D^2 = |\lambda/k_{D'}| v_H^2$. 
Then the scalar potential becomes 
\begin{eqnarray}
	V=(m_{H_u}^2+m_{H_d}^2 + (m_{D'}^2 + m_{\bar{D}'})|\lambda/k_{D'}|)v_H^2.
\end{eqnarray}
Apparently, large $k_{D'}$ leads to an unbounded vacuum, since the contributions from soft masses of $D'$ and 
$\bar{D}'$ are suppressed, and negative $m_{H_u}^2$ dominates over other contributions.
Thus $k_{D'}$ cannot be very large for the stability of the vacuum. Similarly, large $k_{L'}$ is also not allowed. 
Since $m_{L'}^2$ and $m_{\bar{L}'}^2$ are smaller than $m_{D'}^2$ and $m_{\bar{D}'}^2$, 
the upper limit on $k_{L'}$ is more stringent. Therefore the SUSY mass of the extra lepton-like states, $L'$ and $\bar{L}'$, should be as small as $\mathcal O(100 {\rm GeV})$.~\footnote{The breaking of the GUT relation between $k_{L'}$ and $k_{D'}$ may come from the existence of a higher dimensional operator, e.g., $S\, 5(\bar{D}',L') \Sigma(24) \bar{5}/M_{P}$ ($\Sigma(24)$ is the $SU(5)$ GUT breaking Higgs); additional corrections, e.g., $\sim 2 \times 10^{-2} S D' \bar{D}'$ and $\sim -3 \times 10^{-2} S L' \bar{L}'$ may arise.} 
In numerical calculation, we fix $k_{L'}=0.002$. 

%%%%%%%%%%%%%%%%%%%%%%%%%%%%%%%%%
\subsubsection{Stability of the true vacuum}
%%%%%%%%%%%%%%%%%%%%%%%%%%%%%%%%%

There is a potentially dangerous local extremum along $\left< H_u^0 \right> \neq 0$ and $\left< S \right> = \left< H_d^0 \right> = 0$. 
The relative depth of the potential along this direction is estimated as
\begin{eqnarray}
	V_{H_u} = -\frac{2 (m_{H_u}^2)^2}{g_2^2} \sim -10^{13} {\rm GeV}^4.
\end{eqnarray}
for $-m_{H_u}^2 \sim |\mu|^2 \sim 10^6 {\rm GeV}^2$.  
Thus the desired (electroweak symmetry breaking) minimum is deeper : $V_{\rm EWSB} < V_{H_u}$ in the parameter space which we are interested in.

Notice that there is a deeper minimum along the $D$-flat direction for the Higgs:
$|\left<H_u^0 \right>|=|\left<H_d^0 \right>|=v_H$ and $\left<S\right>\sim 0$.
Along this direction, we have the deepest minimum,
\begin{eqnarray}
	V_{H} \simeq -(m_{H_d}^2+m_{H_u}^2)^2/(4\lambda^2) \sim -10^{16}{\rm GeV}^4.
	%V_{H} \simeq \lambda^2 v_H^2 + (m_{H_d}^2 + m_{H_u}^2)v_H^2 ,
\end{eqnarray}
This is deeper than the desired minimum. Thus the desired minimum that we want is metastable and
we need to take care about stability of the desired vacuum.

%\paragraph{Meta Stable Vacuum}
First, let us see the stability of the desired vacuum.
The desired minimum at $v_{H_d}\ll v_{H_u} \ll v_S$ is separated from the global minimum by the potential barrier,
which is of the order of $V_{\rm barrier} \sim \lambda A_\lambda v^3 \sim 10^{15}{\rm GeV}^4$,
where $v\sim 100$\,TeV.
Thus the Euclidean action for the bounce solution connecting the two minima is
estimated to be $S_4 \sim 10V_{\rm barrier} / m_{H_u}^4 \sim 10^4$,
which is sufficient for avoiding the decay of the metastable vacuum into the global minimum
on a cosmological timescale~\cite{Callan:1977pt}.

The next issue to be discussed is whether or not the Universe prefers the desired metastable minimum
rather than the deeper global minimum.
At high temperature regime of $T \gg 1$\,TeV, all particles with masses less than $\sim$TeV
are thermalized and hence they give thermal masses for $S$, $H_u$ and $H_d$.
This stabilizes these fields at the origin : $S=H_u=H_d=0$.
As the temperature decreases, tachyonic directions appear.
Thus the scalars roll down toward the direction which becomes tachyonic first.
By comparing the thermal mass for the $S$ direction, $m_s(T)^2 \simeq 3 N_H k_{D'}^2 T^2 /12$,
and that for the $H_u H_d$ direction, $m_{H_uH_d}(T)^2 \simeq [3(y_t^2+y_b^2) + (15g^2+5g'^2)/2] T^2/12$, 
we obtain the following condition for the $S$ direction to become tachyonic first :
\begin{equation}
	\frac{|m_S^2|}{N_H k_{D'}^2} > \frac{|m_{H_u}^2+m_{H_d}^2|}{(y_t^2+y_b^2) + 5g^2/2+5g'^2/6},
	\label{PT}
\end{equation}
where we have assumed $D'$, top and bottom quarks, gauge bosons, gauginos as well as higgsinos are in thermal bath.
Actually this is satisfied for typical parameters in the model.
Therefore we conclude that the Universe relaxes at the desired minimum toward 
$v_S \sim m_S/\kappa \sim \mathcal O(100)$\,TeV with $v_{H_u}\sim v_{H_d}\sim 0$ at the phase transition.
The $\mu$ and $B\mu$ terms are generated by the VEV of $S$, and the model effectively looks like the MSSM.
After that, Higgs fields found the desired vacuum where $\sqrt{v_{H_u}^2+v_{H_d}^2}=174$\,GeV.
This vacuum is stable against decay into the global minimum, as shown above.\footnote{
	Precisely speaking, the first order phase transition into the global minimum
	may slightly precede the second order one. 
	This does not modify the above result much as long as the phase transition into the true minimum
	occurs much earlier than the second order phase transition into the global minimum; i.e.,
	as long as the condition (\ref{PT}) is well satisfied.
}

%With the convention of $v_u=v \sin\beta$, $v_d=v\cos\beta$,  $v_S > 0$, $\mu>0$ and $\kappa <0$, 
%the true minimum is given by
%%
%\begin{eqnarray}
%	V &=& {m_{H_u}^2}v_u^2 + m_{H_d}^2 v_d^2 + m_S^2 v_S^2 +
%	 \lambda^2 |S|^2 (v_u^2 + v_d^2) + \kappa^2 v_S^4 + \lambda^2 (v_u v_d)^2 \nn
%	&&- (A_\lambda + \kappa v_S) \lambda v_S v_u v_d + h.c. + A_\kappa \kappa v_S^3/3 + {\rm h.c.} ,
%\end{eqnarray}
%%
%where $A_\lambda$ is positive and $A_\lambda -|\kappa|v_S > 0$. Note that we consider the region with large $\tan\beta$, $A_\lambda \simeq |\kappa|v_S$. The deeper minima are found along $\left<S\right>\sim 0$ and $-\left<H_u^0\right> = \left<H_d^0\right> = v_H$ or $\left<H_u^0\right> = \left<H_d^0\right> = v_H$. 
%By substituting $\left<S\right>=(v_S - v)$, the potential can be written as
%\begin{eqnarray}
%V \sim (m_{H_u}^2+m_{H_d}^2+2|\mu|^2+m_S^2)v^2 -4\lambda |\mu| v^3 \pm A_\lambda \lambda v^3 \pm 4\mu\kappa v^3 + \dots 
%\end{eqnarray}
%Therefore the maximum value of the potential can be estimated as
%\begin{eqnarray}
%V_{\rm max} \simeq O(10^{-2})|\mu|^4/\lambda^2,
%\end{eqnarray}
%which is $O(10^{14}) {\rm GeV^4}$ with a typical parameter in our model. 

%%%%%%%%%%%%%%%%%%%%%%%%%%%%%%%%%
\subsection{$B_s \to \mu^+ \mu^-$ and $B \to X_s \mu^+ \mu^-$}\label{B phys}
%%%%%%%%%%%%%%%%%%%%%%%%%%%%%%%%%

When the lightest CP-odd Higgs, $a_1$, is very light, there are sizable contributions to the processes involving B meson~\cite{Hiller:2004ii, Domingo:2007dx, Heng:2008rc,Domingo:2008rr}. 
The flavor violating coupling %, $b$-$s$-$a$ 
among bottom, strange quarks and $a_1$ is generated by non-holomorphic Yukawa couplings, 
which are induced by picking up the CKM matrix element and soft masses at one-loop level~\cite{Isidori:2001fv}.

Since both $\lambda$ and $\kappa$ are small in our model, the mixing between the MSSM-like CP-odd Higgs and the singlet like CP-odd Higgs is as small as $\mathcal O(10^{-4}-10^{-5})$. 
Therefore the contributions to $B_s \to \mu^+ \mu^-$ mediated by $a_1$ are small, resulting in the prediction of 
${\rm Br}(B_s \to \mu^+ \mu^-)\sim \mathcal O(10^{-9})$. 
The branching ratio is about one order of magnitude smaller than the limit from the LHCb result, 
${\rm Br}(B_s \to \mu^+\mu^-) < 1.5 \times 10^{-8} (95\%\, {\rm C.L.})$~\cite{Bettler:2011rp}. \footnote{
{Although, the upper bound for ${\rm Br}(B_s \to \mu^+\mu^-)$ is updated as ${\rm Br}(B_s \to \mu^+\mu^-) < 4.5 \times 10^{-9} (95\%\, {\rm C.L.})$~\cite{current_LHCb}, the most of the parameter space can  satisfy this new bound.}
}
The constraints from $\Delta M_{q}$ $(q=s,d)$ are also loose with such a small mixing~\cite{Domingo:2008rr}.

On the other hand, the inclusive B decay, $B \to X_s \mu^+ \mu^-$ gives a more stringent constraint and excludes some regions of parameter space, even though the mixing is very small. 
When the mass of the lightest CP-odd Higgs is in the ranges of $1 \, {\rm GeV} < M_{\mu^+\mu^-} < 2.4{\rm \, GeV}$ and $3.8 \, {\rm GeV} < M_{\mu^+\mu^-} < 5\,{\rm GeV}$, 
the $a_1$, which mediates this process, can be on-shell. 
Therefore the regions with these CP-odd Higgs masses are excluded in most cases. 
The blue regions shown in Figs.~1-4 are excluded by this constraint.
Thus the regions with small messenger scale and small $\tan\beta$ are not favored. 
Note that the region between 
$2.4\, {\rm GeV} < M_{\mu^+\mu^-} < 3.8\, {\rm GeV}$ can not be constrained, since it is difficult to estimate the branching ratio due to the charm quark resonances. (The corresponding regions are those between two blue strips in Fig.~1-4.)

%%%%%%%%%%%%%%%%%%%%%%%%%%%%%%%%%
\subsection{Small CP violation}
\label{sec:smallCP}
%%%%%%%%%%%%%%%%%%%%%%%%%%%%%%%%%

In NMSSM without CP violating parameters, the CP violating extrema in the potential is not minimal 
but maximal, and hence CP is conserved.
It should be noted, however, that the potential (\ref{eq:Z3b}) 
slightly shifts the position of the minimum of $a_S$ from zero.
Since there is no SUSY CP phase only for $a_S = 0$, the inclusion of this term
induces a small spontaneous CP violation.
The induced CP angle is estimated as
\begin{equation}
	\delta \theta_S = 
	\frac{a_S}{\left<S\right>}
	 \sim \frac{\Lambda_H^4}{m_{a_1}^2 v_S^2}\theta_H  \sim 
	10^{-8} \theta_H \left( \frac{\Lambda_H}{10\,{\rm GeV}} \right)^4
	\left( \frac{10\,{\rm GeV}}{m_{a_1}} \right)^2
	\left( \frac{100\,{\rm TeV}}{v_S} \right)^2.
\end{equation}
The phase of the singlet VEV,  $\delta \theta_S$, gives a phase to that of the effective $\mu$ term, 
$\mu_{\rm eff}=\lambda v_S$, which is constrained by EDM experiment~\cite{Nakamura:2010zzi}. 
Since the upper limit on the CP phase is $\mathcal{O}(10^{-3})-\mathcal{O}(10^{-4})$ for sparticles of 1\,TeV 
and $\tan\beta=\mathcal O(10)$, 
the predicted $\delta \theta_S$ is well below the experimental bound
unless $\Lambda_H$ is much larger than $\mathcal O(10)~{\rm GeV}$.

%%%%%%%%%%%%%%%%%%%%%%%%%%%%%%%%%
\subsection{Hidden gluon and glueball}  \label{sec:hidden}
%%%%%%%%%%%%%%%%%%%%%%%%%%%%%%%%%

The hidden gauge sector contains no light matter.
Thus it is expected that the hidden glueball is the lightest particle in the hidden gauge sector,
and hence it can be long-lived on a collider (and even cosmological) time scale.
Let us see properties of the hidden glueball.

Once a hidden glueball is formed, it can decay into SM particles through loops of
extra matter~\cite{Kang:2008ea,Juknevich:2009gg}.
For example, there are effective dimension 8 operators after integrating out the extra matter,
\begin{equation}
	\mathcal L_{\rm eff} \sim \frac{\alpha_H \alpha'}{m_{\psi_{L'}}^4}\left[ 
		F_{\mu\nu}F^{\mu\nu}{\rm Tr} \left(G_{H \rho\sigma}G_H^{\rho\sigma}\right)
		+ F_{\mu\nu}\tilde F^{\mu\nu}{\rm Tr} \left(G_{H \rho\sigma}\tilde G_H^{\rho\sigma}\right)
	\right],
\end{equation}
where $G_H$ denotes the field strength of the hidden gluon and $\alpha_H$ is the coupling constant
of the hidden gauge symmetry.
The first term induces the scalar glueball decay and the second term the pseudo-scalar glueball decay.
Both have a similar lifetime, which is estimated as
\begin{equation}
	\tau (g_H \to \gamma\gamma) \sim 6\times 10^{-5}{\rm sec} \left( \frac{10\,{\rm GeV}}{\Lambda_H} \right)^9
	\left( \frac{m_{\psi_{L'}}}{1\,{\rm TeV}} \right)^8,  \label{life8}
\end{equation}
where we have simply approximated all the mass scales 
appearing in the pure hidden gluon sector, the glueball mass and its decay constant, to be $\Lambda_H$. 
Thus Eq.~(\ref{life8}) should be regarded as only an order-of-magnitude estimation,
but it is sufficient to see that the lifetime can be short enough to be
free from the BBN constraint, and long enough to pass through the detector at collider experiments
without leaving signals except for missing energies.
Similarly, they can decay into SM gluons via dimension 8 operators, but in this case
it is $D'$ that mediates the decay, and it is heavier than $L'$.
Due to the high powers of $m_{D'}$ in the expression of lifetime, 
it may not be the dominant decay mode.

Moreover, the one-loop coupling of hidden gluons to the singlet scalar leads
to the following effective dimension 7 operators
\begin{equation}
	\mathcal L_{\rm eff}\sim \frac{\alpha_H}{8\pi v_S}
	\left[ 
		\frac{\epsilon_{s h}}{m_S^2-\Lambda_H^2}\bar f f{\rm Tr} \left(G_{H \rho\sigma}G_H^{\rho\sigma}\right)
		+ \frac{\epsilon_{a_S a_h}}{m_{a}^2-\Lambda_H^2}
		\bar f\gamma_5 f {\rm Tr} \left(G_{H \rho\sigma}\tilde G_H^{\rho\sigma}\right)
	\right],
\end{equation}
where $f$ denotes SM fermions which are mediated by exchange of scalar and pseudo-scalar Higgs bosons.
Here $\epsilon_{s h}$ and $\epsilon_{a_S a_h}$ represent mixings between
the singlet (pseudo) scalar bosons and SM (pseudo) scalar Higgs bosons.
For example, the lifetime of the scalar glueball into the charm pair is estimated as
\begin{equation}
	\tau(g_H \to c\bar c) \sim 5\times 10^{-4}{\rm sec}
	\left( \frac{10\,{\rm GeV}}{\Lambda_H} \right)^7
	\left( \frac{v_S}{100\,{\rm TeV}} \right)^2
	\left( \frac{m_S}{100\,{\rm GeV}} \right)^4
	\left( \frac{0.1}{\epsilon_{sh}} \right)^2, \label{life8b}
\end{equation}
and similar expression holds for the pseudo scalar glueball.
This is close to that induced by dimension 8 operators (\ref{life8}),
but the parameter dependences are different.
Taking account of uncertainties in these lifetime estimates coming from the strong hidden gauge sector,
we do not regard these estimates to be robust, but we mention 
possible cosmological effects of hidden glueballs if the lifetime happens to be close to 1 sec.
In any case, it is reasonable to suppose that the dynamical scale of the hidden gauge group should satisfy
\begin{equation}
	\Lambda_H \gsim \mathcal O (1)~{\rm GeV},
\end{equation}
in order for the glueball lifetime to be shorter than 1 sec.
Note that there exist many excited states of hidden glueballs
but they decay into a ground states of scalar or pseudo scalar glueballs
with lifetime comparable or shorter than the above estimate~\cite{Juknevich:2009gg}.

%%%%%%%%%%%%%%%%%%%%%%%%%%%%%%%%%
\subsection{Hidden gaugino}   \label{sec:gaugino}
%%%%%%%%%%%%%%%%%%%%%%%%%%%%%%%%%

\begin{figure}[tbp]
\begin{center}
 \includegraphics[scale=1.0]{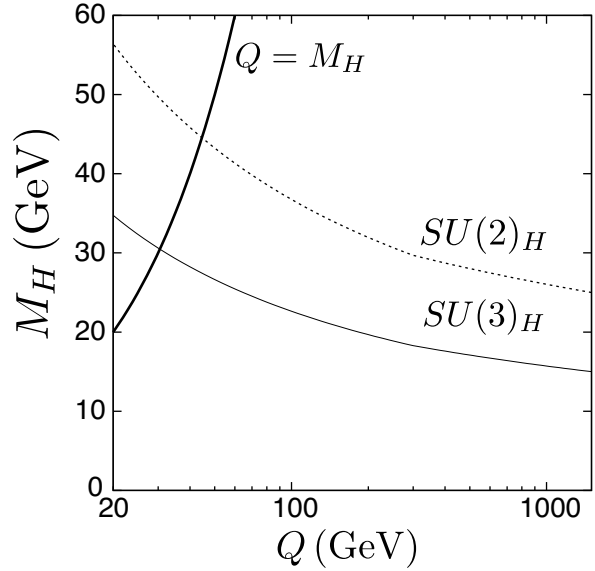}
\caption{
The running masses of hidden gauginos are shown. The thin solid line corresponds to $\SU(3)_H$ case, while the dotted line corresponds to the case of $\SU(2)_H$. The thick solid line represents $Q=M_H$, where $Q$ is a renormalization scale.
%The green line shows $Q=M_{H}$, where $Q$ is renormalization scale.
The gauge coupling of $\SU(N_H)$ is taken to be $g_H(Q=1.5{\rm TeV})=1.392$ for $N_H=2$ and 1.122 for $N_H=3$, which corresponds $\Lambda_H\sim 5$ GeV.
}
\label{fig:m_gaugino}
\end{center}
\end{figure}

The mass of hidden gauginos are dominantly generated by two-loop RGE running effects (see Appendix A), 
and they are 15-25 GeV at the SUSY scale ($\sim$1.5 TeV). 
However, the hidden gaugino mass becomes large when the renormalization scale becomes lower. 
At the one-loop level the following relation holds, 
\begin{eqnarray}
\frac{M_H (m_{\psi_{L'}})}{g_{H}^2(m_{\psi_{L'}})} = \frac{M_H(M_H)}{g_{H}^2(M_H)}.
\end{eqnarray}
If we demand $\Lambda_H \sim 5\,{\rm GeV}$, the running mass of 
the hidden gaugino becomes about 20-50 GeV. 
Typical values of the running hidden gaugino masses are shown in Fig.~{\ref{fig:m_gaugino}}.
In the case where $M_H < M_Z/2$ is satisfied, 
the $Z$ boson can decay into hidden gauginos at the one-loop level. 
However, the branching ratio is sufficiently small, and the constraint from the invisible
$Z$ decay width~\cite{:2005ema} is easily avoided.

%%%%%%%%%%%%%%%

%%%%%%%%%%%%%%%%%%%%%%%%%%%%%%%%%
\subsection{Extra Matter}
%%%%%%%%%%%%%%%%%%%%%%%%%%%%%%%%%

The typical mass of the extra lepton, $m_{\psi_{L'}}$, is 200-400 GeV with $k_{L'}=0.002$ (see. Table.\ref{tab:mass}). 
Since a large value of $k_{L'}$ is not allowed by the constraints on the vacuum structure (see \ref{CCB}), 
the mass of the extra leptons cannot be as large as 1\,TeV. 
On the other hand, the mass of the extra-quark, $m_{\psi_{D'}}$, is predicted to be 3-4.5 TeV. 
The size of $k_{D'}$, and hence $m_{\psi_{D'}}$ is limited by the requirement for
the successful EWSB and non-existence of unbounded vacuum. 

As for the scalar states, there are additional contributions from soft SUSY breaking masses generated by gauge mediation effects. 
Since the SM charges of $\bar{D'}$ are the same as those of down-type (s)quark, the soft breaking mass $(m_{\tilde{D'}})$ is predicted to be $\sim 1.5$ TeV. 
Similarly, the soft mass of the extra slepton, $m_{\tilde{L'}}$, is almost the same as that of $SU(2)_L$ sleptons. 

In most of the viable parameter regions, the lighter mass eigenstate of the extra squark, $\tilde{D}'_1$ is a bit lighter than its fermionic partner, $D'$, due to the mixing induced by $A_{D'}$. 
The extra sleptons are heavier than the extra-leptons, and their masses are about 500 GeV.

%%%%%%%%%%%%%%%
%\begin{figure}[tbp]
%\begin{center}
%\includegraphics[scale=1.0]{coupling_ratio.eps}
%\caption{
%	The ratio of $\alpha_3/\alpha_2$ are shown. Due to the larger ratio of the coupling in $N_H=3$, the mass of the sleptons and squarks tends to be split, which is preferable in terms of the radiative corrections to Higgs mass and muon g-2.
%}
%\label{fig:ratios}
%\end{center}
%\end{figure}
%%%%%%%%%%%%%%%

%%%%%%%%%%%%%%%%%%%%%%%%%%%%%%%%%%%%%%%%%%%%%
\section{Cosmological issues}     \label{sec:cos}
%%%%%%%%%%%%%%%%%%%%%%%%%%%%%%%%%%%%%%%%%%%%%

In the present model there are lots of cosmological issues to be discussed.
Because of the hidden gauge symmetry, there appear stable or long-lived particles
which may potentially spoil success of the standard cosmology.
We address these issues in this section.

%%%%%%%%%%%%%%%%%%%%%%%%%%%%%%%%%%%%%%%%%%%%%
\subsection{Singlino}
\label{sec:singlino}
%%%%%%%%%%%%%%%%%%%%%%%%%%%%%%%%%%%%%%%%%%%%%

The lightest SUSY particle in the NMSSM sector, except for the gravitino,
is the lightest neutralino, which mostly consists of the singlino ($\tilde s$) 
whose mass is around 60\,GeV for typical interesting parameters.
On the other hand , the hidden gaugino ($\tilde g_H$) also has a mass of around 40\,GeV for typical parameters
(see Sec.~\ref{sec:gaugino}).
Cosmological scenarios significantly change depending on which is the lighter.
We discuss cosmology of each case separately.

First, let us assume that the singlino is lighter than the hidden gaugino.
The main channel of the singlino annihilation is into 
singlet higgs bosons, $a_1$ and $h_1$, both of which mostly consist of singlet.
However, the annihilation cross section is significantly suppressed due to the smallness of the
coupling $\kappa (\sim 10^{-4})$ in our model (cf.~\cite{Kappl:2010qx}).
Therefore the singlino decouples from thermal bath soon after the temperature becomes lower than the mass of the hidden gauginos.
Thus the singlino abundance is huge and it may even dominate the Universe soon after the freezeout.
The singlino decays into the gravitino and $a_1$ with a lifetime of
\begin{equation}
	\Gamma(\tilde{s} \to a_1 + \psi_{3/2})^{-1} \simeq
	1.9\times 10^{-2}~{\rm sec}
	\left( \frac{50\,{\rm GeV}}{m_{\tilde s}} \right)^5
	\left( \frac{m_{3/2}}{0.1\,{\rm MeV}} \right)^2. \label{eq:g_width}
\end{equation}
If the singlino dominantly decays into the gravitino, a large amount of the gravitino is produced. 
%Its energy density becomes as large as 
%\begin{eqnarray}
%	\Omega_{3/2} h^2 \sim X \left(\frac{m_{3/2}}{0.1 {\rm MeV}}\right) 5.6 \times 10^2,
%\end{eqnarray}
%with typical parameters. 
Apparently such large energy density conflicts with the WMAP observation, 
even if there exists a dilution process due to the late-time entropy production.
Thus it is difficult to reproduce the standard BBN in this setup.

 Next, let us consider the opposite case : the hidden gaugino is lighter than the singlino.
This is actually the case in the model points P1, P2 and P3 in Table~\ref{tab:mass}.
In this case, the above csomological difficulty can be avoided because
the singlino can decay into hidden gluino and hidden gluon through the dimension 5 operator:\footnote{
	We neglect the SUSY breaking effects, which does not change the result significantly.
}
\begin{eqnarray}
	\mathcal{L} = \int d^2 \theta \frac{\alpha_H}{4\sqrt{2}\pi (v_S / N)} SW^{(a)\alpha}W^{(a)}_\alpha + h.c.
\end{eqnarray}
where $\alpha_H=g_H^2/4\pi$, $N=5$ is the number of vector-like matter fields, 
and $W^{(a)}$ is the gauge supermultiplet of the hidden group.
The lifetime is given by
\begin{equation}
	\Gamma(\tilde{s} \to g_H + \tilde{g}_H)^{-1} \simeq
	10^{-17}~{\rm sec}
	\left(\frac{1}{\alpha_H^2 (N_H^2-1)}\right)
	\left( \frac{50\,{\rm GeV}}{m_{\tilde s}} \right)^3
	\left( \frac{v_S}{100\,{\rm TeV}} \right)^2
	\left( 1- \frac{m_{{\tilde g}_H}^2}{m_{\tilde s}^2} \right)^{-3}. 
\end{equation}
%%
%Therefore the singlino dominantly decay into the hidden gluino and gluon.
The singlino lifetime is much shorter than the cosmological time scale of the hidden gaugino decoupling.
Thus only the hidden gaugino is left after the freezeout
whose abundance is sufficiently small and decays into gravitino before BBN begins
as described Sec.~{\ref{sec:hglu}}.
 There is no cosmological difficulty associated with the singlino in this case.
Hereafter we only consider this case.

%%%%%%%%%%%%%%%%%%%%%%%%%%%%%%%%%%%%%%%%%%%%%
\subsection{Hidden glueball and gaugino}\label{sec:hglu}
%%%%%%%%%%%%%%%%%%%%%%%%%%%%%%%%%%%%%%%%%%%%%

Next, we discuss cosmological consequences of the hidden gauge sector.
After the phase transition occurring at $T\sim \Lambda_H$, hidden gluons form color singlet states,
so-called hidden glueballs.
As shown in Eqs.~(\ref{life8})(\ref{life8b}), the lifetime of the hidden glueball can be short enough to be
free from the BBN constraint but can also be long enough to modify thermal history before BBN.
Hereafter we take the typical scale of the hidden QCD to be $\Lambda_H \sim 10$\,GeV.
If it is much smaller, the glueball lifetime becomes too long. If it is much larger,
the induced CP angle causes a too large EDM (see Sec.~\ref{sec:g-2}).

Glueballs are formed at the temperature $T\sim \Lambda_H$.
Hidden gluons keep thermal equilibrium with the SM sector until $D'$ and $L'$ are decoupled
at $T\sim m_{\psi_{L'}}/20$. This is not far from $\Lambda_H$, and hence we assume 
hidden gluons were in thermal equilibrium at the formation of glueballs,
although this assumption does not affect the following result at all.
At the formation, the energy density of thermal plasma is around $\sim (\pi^2 g_*/30) \Lambda_H^4$
and that of the glueball is around $\sim m_{g_H} \Lambda_H^3$
with $m_{g_H}$ being the glueball mass.
Since $m_{g_H} \gtrsim \Lambda_H$, the hidden glueballs soon begin to dominate the Universe.
Its lifetime was estimated in Sec.~\ref{sec:hidden}.
There we have shown that the lifetime can be shorter than $0.1$\,sec for $\Lambda_H \gtrsim \mathcal O(1)$\,GeV.
If the lifetime of the glueball is shorter than $10^{-8}$\,sec, 
which corresponds to a temperature above 10\,GeV,
it decays as soon as it is formed, and has no effects on cosmology.
On the other hand, if the lifetime is much longer than $10^{-8}$\,sec but shorter than 0.1\,sec, 
the hidden glueball dominates the Universe until it decays at $T = T_d$.
In this case, the number density of all relics 
existing before the phase transition, such as the gravitino and extra matter to be discussed
as well as the baryon asymmetry,
are diluted by the factor $\sim (T_d/\Lambda_H)$.
We will comment on cosmological effects of late time hidden glueball decay in Sec.~\ref{sec:scenario}.

The hidden gaugino $(\tilde g_H)$ effectively annihilates into hidden gluon pair at the freezeout
and the abundance is significantly reduced.
After the phase transition, a hidden gaugino may form a bound state 
with a hidden gluon and such a bound state may effectively find a partner 
due to its large geometrical cross section. Then the abundance may be further reduced.
In any case, the hidden gaugino eventually decays into the gravitino $\psi_{3/2}$ and the hidden gluon.
The lifetime is estimated as
\begin{equation}
	\tau(\tilde g_H \to g_H + \psi_{3/2}) \simeq 
	1.9\times 10^{-2}~{\rm sec}\left( \frac{50\,{\rm GeV}}{m_{\tilde g_H}} \right)^5
	\left( \frac{m_{3/2}}{0.1\,{\rm MeV}} \right)^2,
\end{equation}
The gravitino abundance produced by the decay of hidden gluino is sufficiently small.
Therefore, it does not have significant cosmological effects.

%%%%%%%%%%%%%%%%%%%%%%%%%%%%%%%%%%%%%%%%%%%%%
\subsection{Extra matter}
%%%%%%%%%%%%%%%%%%%%%%%%%%%%%%%%%%%%%%%%%%%%%

The lightest particle in the sector of extra matter is stable due to the conservation of the 
hidden gauge charge. 
The colored particle, $D'$, is heavier than the lepton-like one, $L'$, since $k_{D'} > k_{L'}$.
Scalar components obtain SUSY breaking effects from gauge mediation.
The typical mass relation here is $m_{\tilde{D}'_{2}} >  m_{\psi_{D'}} >  m_{\tilde{D}'_{1}} > m_{\tilde{L}'_{1,2}} > m_{\psi_{L'}}$,
where $m_{\tilde{D}'_1}$ and $m_{\tilde{D}'_2}$ ($m_{\tilde{L}'_1}$ and $m_{\tilde{L}'_2}$)
are mass eigenvalues of scalar components of $D'$ and $\bar D'$ ( $L'$ and $\bar L'$).
We discuss fate of these particles.

Here we make a brief comment on the splitting of the electrically neutral and charged component of $L'$.
The electrically charged component of $L'$, denoted by $L^{-'}$, is slightly heavier than
its neutral component, $L^{0'}$, due to  electroweak radiative corrections, by a typical amount
of $\sim 100$MeV~\cite{Thomas:1998wy, Cirelli:2009uv}. 
Thus $L^{-'}$ can decay into $L^{0'}$ and $e \nu$ with a lifetime much shorter than
the cosmological timescale.
Moreover, its scalar partner, $\tilde L^{0'}$, decays into $L^{0'}$ by gaugino exchange
if $\tilde L^{0'}$ is not the lightest $R$-odd particle.
Since this decay process occurs very quickly, the electrically charged component of $L'$
is of no importance in the following discussion.
Similarly, for the $D'$ multiplet, the lighter scalar component $\tilde D'_1$ is left among the supermutiplet.

Now let us follow the evolution of extra matter, $L^{0'}$ and $D'$. 
(For notational simplicity, we write $L^{0'}$ as $L'$ and $\tilde D_1'$ as $D'$ hereafter.)
It effectively annihilates into hidden gluon pairs at the freeze out.
By using the cross section of $\sim \pi\alpha_H^2 / m_{\psi_{L'}}^2$, the resulting abundance is
estimated as
\begin{equation}
	Y_{L'} \equiv \frac{n_{L'}}{s} \simeq 4\times 10^{-15}\left( \frac{m_{\psi_{L'}}}{1\,{\rm TeV}} \right).
\end{equation}
A similar expression holds for $D'$.
Soon after the freezeout, the hidden QCD phase transition takes place at $T\sim \Lambda_H$,
and then these relic $L'$ and $D'$ particles are connected by strings of hidden gauge fluxes.
Since there are no other light matter which can form pair with $L'$ and $D'$, 
this matter-string system is stable against creations of $L'$-$\bar L'$ or $D'$-$\bar D'$ pairs.
Thus we need to examine whether this kind of system eventually annihilates or not.

First, we study the dynamics of $L'$-$\bar L'$ pair connected by strings.
The mean separation $\ell$ among $L'$ particles at the epoch of formation is estimated to be
$\ell \sim n_{L'}^{-1/3}\sim Y_{L'}^{-1/3} T^{-1} \sim 10^5 \Lambda_H^{-1}$. 
The energy stored in one string is thus $E \sim \Lambda_H^2 \ell_{\rm str} \sim 10^5 \Lambda_H$,
which is considerably larger than $m_{\psi_{L'}}$, 
with $\ell_{\rm str} \sim \ell$ being the initial typical string length.\footnote{
	Initial correlation length of the string $(\xi)$ is expected to be of order of $\xi \sim \Lambda_H^{-1}$.
	This is much shorter than the mean separation of $L'$ particles.
	Thus, the string connecting them have Brownian motion like structure with each step size of $\xi$.
	Then the string may have length of order of $\ell_{\rm str} \sim \ell^2 / \xi \sim 10^{10}\Lambda_H^{-1}$
	right after the formation.
	It might be expected that these small scale irregularities are soon smoothed out due to the string tension
	and string self-interactions, and eventually a straight string connecting $L'$-$\bar L'$ pair is left.
	Even if we simply regard $\ell_{\rm str}$ as a maximum and typical initial string length, following 
	energy loss mechanisms will dissipate the string energy effectively and the discussion is not affected.
}
The energy density of the hidden string network is estimated as 
$\rho_{\rm str} \sim \Lambda_H^2\ell_{\rm str} \ell^{-3} \sim 10^{-10}\Lambda_H^4$, which is much smaller
than the total energy density of the Universe.
String forces attract the $L'$-$\bar L'$ pair, and it easily accelerates the $L'$ to the energy of $E (\gg m_{\psi_{L'}})$.
In order for it to be annihilated, it must lose its energy so that it can form a bound state with $\bar L'$.
There exist several energy loss processes of the matter-string systems during the oscillation :
(i) hidden gluon emission and (ii) scattering with background plasma. Let us see them in detail.\footnote{
	Gravitational radiations from moving string system also act as an energy loss mechanism,
	but the timescale of gravitational energy loss is much longer than the following processes.
	See e.g., Ref.~\cite{BlancoPillado:1999cy}.
}\\

(i) Hidden gluon emission \\

One of the possible energy loss processes is emissions of hidden gluons.
Since the initial energy $E$, carried by $L'$ and $\bar L'$, is much larger than the hidden glueball mass,
hidden gluon emissions may take place.
Note, however, that the acceleration by the string tension is around $\sim \Lambda_H^2/m_{\psi_{L'}}$,
which is much smaller than the hidden glueball mass $\sim \Lambda_H$.
Thus emissions occur only when $L'$ and $\bar L'$ meet together on separation of $\sim \Lambda_H^{-1}$.
As a rough but reasonable estimate, let us assume one hidden glueball is emitted when
a $L'$-$\bar L'$ pair passes through each other during one oscillation~\cite{Kang:2008ea,Harnik:2011mv}.\footnote{
	Here we assume that the string connects the $L'$-$\bar L'$ pair in an almost straight line. 
	Otherwise, the probability that $L'$ finds $\bar L'$ in one oscillation is significantly suppressed.
}
The oscillation period is given by,
\begin{equation}
	t_{\rm osc}\sim \left \{ \begin{array}{ll}
	\sqrt{m_{\psi_{L'}}\ell /\Lambda_H^2} & ~~{\rm for}~~E=\Lambda_H^2 \ell < m_{\psi_{L'}} \\
	\ell					   & ~~{\rm for}~~E=\Lambda_H^2 \ell > m_{\psi_{L'}} \\
	\end{array}
	\right.
\end{equation}
and here we conservatively assume the emitted hidden glueball energy is $\sim \Lambda_H$.
The energy loss rate is calculated as $dE/dt \sim \Lambda_H/t_{\rm osc}$.
The typical energy loss time is compared to the Hubble time as
\begin{equation}
	\frac{E/(dE/dt)}{t_{\rm Hub}} \sim \frac{T^2 E^{3/2}m_{\psi_{L'}}^{1/2}}{M_P \Lambda_H^3}
	\sim 10^{-13}\left( \frac{T}{10\,{\rm GeV}} \right)^2
	\left( \frac{E}{1\,{\rm TeV}} \right)^{3/2}
	\left( \frac{m_{\psi_{L'}}}{1\,{\rm TeV}} \right)^{1/2}
	\left( \frac{10\,{\rm GeV}}{\Lambda_H} \right)^3.
	\label{tloss_hid1}
\end{equation}
for $E < m_{\psi_{L'}}$, and
\begin{equation}
	\frac{E/(dE/dt)}{t_{\rm Hub}} \sim \frac{T^2 E^{2}}{M_P \Lambda_H^3}
	\sim 10^{-13}\left( \frac{T}{10\,{\rm GeV}} \right)^2
	\left( \frac{E}{1\,{\rm TeV}} \right)^{2}
	\left( \frac{10\,{\rm GeV}}{\Lambda_H} \right)^3.
	\label{tloss_hid2}
\end{equation}
for $E > m_{\psi_{L'}}$.
Therefore, the hidden gluon emission effectively dissipates energies of strings, 
if the strings soon relax to a rod-like structure connecting $L'$-$\bar L'$ pair after the formation.\\

(ii) Scattering with background particles \\

We here note that it is not clear whether a string-matter system relaxes to a rod-like structure 
so that $L'$ and $\bar L'$ pass through each other at each oscillation, due to a lack of understandings 
of the dynamics such strings in a cosmological setup.
Even if the hidden gluon emission does not work, however,
there are another process to reduce the energy of the system.
Since $L'$ ($\bar L'$) has a weak interaction,
it scatters off the background plasma consisting of leptons and light quarks
through tree level interactions.
Let us consider the interaction of $L'$ with background plasma through the $Z$-boson exchange, 
when $L'$ obtains a large energy $E$ due to the string tension.
The typical scattering cross section is given by $\langle\sigma v \rangle\sim \alpha^2/(ET)$
for $ET > m_{\psi_{L'}}^2$ and it loses its energy of order of $E$ in one collision. 
Therefore, the energy loss rate is calculated as $dE/dt = n_T \langle\sigma v \rangle \Delta E$
where $n_T \sim T^3$ denotes the number density of background plasma.
By comparing the energy loss time scale with the Hubble time, we obtain
\begin{equation}
	\frac{E/(dE/dt)}{t_{\rm Hub}} \sim \frac{E}{\alpha^2 M_P}
	\sim 10^{-13}\left( \frac{E}{1\,{\rm TeV}} \right).
\end{equation}
Thus the energy of matter-string system is quickly reduced.
After the energy decreases to $ET < m_{\psi_{L'}}^2$,  
the typical scattering cross section is given by $\langle\sigma v \rangle\sim G_F^2(ET/m_{\psi_{L'}})^2$
and an average energy loss per scattering is given by $\Delta E \sim TE^2 / m_{\psi_{L'}}^2$.
By comparing the energy loss time scale with the Hubble time, we obtain
\begin{equation}
	\frac{E/(dE/dt)}{t_{\rm Hub}} \sim \frac{m_{\psi_{L'}}^4}{M_P G_F^2 E^3 T^4}
	\sim 10^{-8}\left( \frac{1\,{\rm TeV}}{E} \right)^3\left( \frac{10\,{\rm GeV}}{T} \right)^4
	\left( \frac{m_{\psi_{L'}}}{1\,{\rm TeV}} \right)^4.
\end{equation}
Thus the $L'$ would soon be non-relativistic well within one Hubble time.
After it becomes non-relativistic,
the scattering cross section is given by $\langle\sigma v \rangle\sim G_F^2 T^2$
and an average energy loss per scattering is $\Delta E \sim T E/m_{\psi_{L'}}$ for $E < m_{\psi_{L'}}$.
Here $E$ is regarded as the kinetic energy of $L'$.
Then we obtain
\begin{equation}
	\frac{E/(dE/dt)}{t_{\rm Hub}} \sim \frac{m_{L'}}{M_P G_F^2 T^4}
	\sim 10^{-8}\left( \frac{10\,{\rm GeV}}{T} \right)^4 \left( \frac{m_{\psi_{L'}}}{1\,{\rm TeV}} \right).
\end{equation}
Therefore, we conclude that scatterings of $L'$ particles with thermal plasma efficiently dissipate the
string energy well within one Hubble time. The resultant energy of the matter-string system,
or the kinetic energy of $L' (\bar L')$, is expected to be of order of $T$.
Since this kinetic energy is comparable or even smaller than $\alpha_H^2 m_{\psi_{L'}}$,
we reasonably expect the formation of $L'$-$\bar L'$ bound states.
Once the bound state forms, the annihilation takes place efficiently 
as shown in Refs.~\cite{Kang:2008ea,Fok:2011yc}.
After all, all the extra matter disappear after the hidden QCD phase transition.

Almost the same arguments hold for $D'$-$\bar D'$ pair connected by strings.
Since $D'$ has color and electric charges, its energy loss is expected to be much more efficient than
the $L'$ studied above.
Thus $D'$-$\bar D'$ pair also annihilates soon after the hidden QCD phase transition.

A complexity arises for $D'$-$L'$ and $\bar D'$-$\bar L'$ pairs connected by strings.
After they lose energies by the same processes described above, they form bound states.
They are stable because of the conservation of the hidden baryon (lepton) numbers.
Since it has a QCD color charge, it further forms a bound state with $\bar d (d)$ after the QCD phase transition.
The bound states $D' L' \bar d$ and $\bar D' \bar L' d$ find each other with the cross section of 
$\Lambda_{\rm QCD}^{-2}$.
After the collision, the system loses energy by the photon emission and finally annihilates rapidly
\cite{Kang:2006yd}.
Eventually, the relic abundance of $D' L' \bar d$ and $\bar D' \bar L' d$ bound states is given by~\cite{Kang:2006yd}
\begin{equation}
	Y_{\rm bound~state} \simeq 10^{-18}\left( \frac{m_{\psi_{D'}}}{1\,{\rm TeV}} \right)^{1/2} ,
\end{equation}
This is much smaller than the dark matter abundance, but it is still subject to a constraint from BBN
\cite{Kusakabe:2009jt,Kawasaki:2010yh}.
Taking account of the dilution by the entropy production, 
this may be marginally allowed.
It is also possible to avoid the constraint by introducing the following operator\footnote{
	This operator can be consistent with the $Z_3$ symmetry as well as the R-parity, 
	with a slightly modified charge assignment: 
	R$_p (D')$ = R$_p(\bar D') = +$, $Z_3(\bar D') = Z_3(\bar L') = 0$, and $Z_3(D') = Z_3(L') = 2$.
	Note that this change does not affect the discussion so far at all.
}
\begin{equation}
	W = \frac{1}{M}D' \bar d L' H_u,
\end{equation}
with $M$ being the cutoff scale, which may be expected to be $\mathcal O(10^{16})$\,GeV.
This induces the decay of $D'$ into $L'$ with lifetime of $\mathcal O(1)$\,sec.
 After the decay, $L'$-$\bar L'$ bound states connected by strings are left
with $L'$ particles initially having large kinetic energies. 
Although the energy loss through scatterings with thermal plasma is inefficient at this epoch,
they soon lose energies through hidden gluon emissions~(See Eqs.~(\ref{tloss_hid1}) and (\ref{tloss_hid2})), 
since the strings are straight and they pass through each other in each oscillation.
Therefore, after the $D'$ decay, the bound states soon disappear.

%%%%%%%%%%%%%%%%%%%%%%%%%%%%%%%%%%%%%%%%%%%%%
\subsection{Cosmological scenario}   \label{sec:scenario}
%%%%%%%%%%%%%%%%%%%%%%%%%%%%%%%%%%%%%%%%%%%%%

Now let us summarize the cosmological scenarios in our model.
As we have discussed, 
the lifetime of hidden glueball is very sensitive to the model parameters.
%Depending on its lifetime, two extreme scenarios can be considered.
In the short lifetime limit, hidden glueballs do not have significant cosmological effects
and we recover standard thermal history after reheating.
In the long lifetime limit, glueballs once dominate the Universe and they finally decay 
slightly before BBN begins.
In this case a significant amount of entropy is released.
We discuss cosmological scenarios with and without entropy production,
although the latter is likely to be realized.
Let us repeat here that domain walls that are formed after electroweak phase transition 
decay rapidly and harmless in our model due to the bias induced by the hidden QCD instanton effect.
 The singlino does not significantly affect cosmology if it is heavier than the hidden gaugino. 
Extra matter particles also quickly disappear after the hidden QCD phase transition
due to the dynamics of strings connecting them.

%%%%%%%%%%%%%%%%%%%%%%%%%%%%%%%%%%%%%%%%%%%%%
\subsubsection{Without entropy production}
%%%%%%%%%%%%%%%%%%%%%%%%%%%%%%%%%%%%%%%%%%%%%

First of all, we consider the scenario without late-time entropy production;
that is, the hidden glueball lifetime is rather short.
The reheating temperature, $T_{\rm R}$, is bounded above from the gravitino overproduction as
$T_{\rm R}\lesssim 10^3$\,GeV for $m_{3/2}=0.1$\,MeV~\cite{Moroi:1993mb,Bolz:2000fu,Pradler:2006qh,Rychkov:2007uq,Cheung:2011nn}.
The gravitino can be the main component of dark matter if the reheating temperature is close to this upper bound.

In the presence the PQ sector, the axino ($\tilde a$), a fermionic superpartner of the axion, is also produced thermally.
The abundance is estimated as~\cite{Covi:2001nw,Brandenburg:2004du,Strumia:2010aa,Choi:2011yf} 
\begin{equation}
	Y_{\tilde a}\equiv \frac{n_{\tilde a}}{s}\sim 2\times 10^{-5}
	\left( \frac{T_{\rm R}}{10^6\,{\rm GeV}} \right)
	\left( \frac{10^{11}\,{\rm GeV}}{f_a} \right)^2.
\end{equation}
This expression is valid also for $T_R\gtrsim 1$\,TeV.
The axino mass sensitively depends on the axion model,
but it can be as heavy as the gravitino~\cite{Chun:1995hc}.
The axino overproduction is avoided for $f_a \gtrsim 10^{11}$\,GeV
if the axino mass is comparable to the gravitino mass.\footnote{
	If the axino is lighter than the gravitino, the gravitino can decay into axino and axion,
	but its lifetime is longer than the present age of the Universe.
	Thus the bound on the $T_{\rm R}$ does not change.
}
The saxion ($\sigma$), the scalar partner of the axion, is also produced
through thermal scatterings and coherent oscillation.
It generally obtains a mass of at least of the gravitino, and decays into two axions with lifetime
$\tau \sim 1\times 10^{12}{\rm sec}(0.1{\rm MeV}/m_\sigma)^3(f_a/10^{11}{\rm GeV})^2$.
This decay process is constrained from the observation of effective number of neutrino species
in the cosmic microwave background anisotropy.
The reheating temperature of $T_{\rm R}\sim 10^3$\,GeV is marginally consistent with
this constraint (see e.g., Ref.~\cite{Kawasaki:2007mk}).
But the saxion may obtain much larger mass than the gravitino in gauge-mediation~\cite{Asaka:1998ns,Banks:2002sd}. Then the lifetime becomes much shorter and the constraint is significantly relaxed.
The cosmology of saxion depends much on the model of PQ sector, particularly, on how the PQ scalar
is stabilized at the scale of $f_a$. Here it is sufficient to note that cosmological constraints on the saxion
can rather easily be avoided.
Finally, for the PQ scale of $f_a \sim 10^{11}$\,GeV, the axion coherent oscillation
is another good dark matter candidate.
Thus in this case the dark matter may be a mixture of thermally produced gravitino and the axion.

%%%%%%%%%%%%%%%%%%%%%%%%%%%%%%%%%%%%%%%%%%%%%
\subsubsection{With entropy production}
%%%%%%%%%%%%%%%%%%%%%%%%%%%%%%%%%%%%%%%%%%%%%

A more interesting cosmological scenario appears in the case of late-time entropy production from 
the hidden glueball decay.
Here we assume the lifetime of hidden glueball is around $0.1$\,sec so that their final decay temperature
of them is around a few MeV.
In this case the entropy production
dilutes all the abundance of all relics existing before the hidden QCD phase transition 
by up to an amount of $\Delta \sim 10^3$.
Under this circumstance, the reheating temperature after inflation can be raised
up to, say, $T_{\rm R}\sim 10^{10}$\,GeV~\cite{Fujii:2003iw}.
This is because light gravitinos are thermalized for such a high reheating temperature 
and their abundance is determined by thermal one, $Y_{3/2} = 0.417/g_{*3/2}$
where $g_{*3/2}$ denotes the relativistic degrees of freedom at the decoupling of the gravitino.\footnote{
	Note that $g_* \simeq 370$ in our model if all the matter and gauge fields are thermalized
	(except for the messenger and SUSY breaking sector).
}
Translating into the density parameter, we have 
$\Omega_{3/2}h^2 = 0.3(m_{3/2}/0.1{\rm MeV})(300/g_{*3/2})(10^2/\Delta)$.
An attractive feature is that in such a case, (non-)thermal leptogenesis scenarios
\cite{Fukugita:1986hr,Asaka:1999yd,Asaka:1999jb}
become viable even under the entropy production~\cite{Fujii:2003iw}.

Note, however, that if the reheating temperature exceeds the messenger scale,
the messenger fermions are also efficiently produced and take part in thermal bath.
This is problematic because the lightest messengers are stable due to a messenger parity.\footnote{
Also the messenger coupling to the goldstino can enhance the gravitino production~\cite{Choi:1999xm}.
}
One can avoid this by demanding $T_{\rm R} \lesssim 10^9$\,GeV,
where nonthermal leptogenesis still works under the entropy production.
It is also possible to break the messenger parity to a small amount so that messengers can decay.
For example, one is allowed to introduce the following terms\footnote{
	Here the $Z_3$ charges of $\Psi_{D,\bar{L}}$ and $\Psi_{\bar{D},{L}}$ are slightly modified as
	$Z_3(\Psi_{D,\bar{L}})=1$ and $Z_3(\Psi_{\bar D,L})=2$, respectively.
}
\begin{eqnarray}
	W = g_{D,i} S \bar{d}_i \Psi_{D} + g_{L,i} S {\ell}_i \Psi_{\bar{L}},
\end{eqnarray}
with tiny coupling constants $g_{D,i}$ and $g_{L,i}$, and $\bar d_i$ and $\ell_i$ are the MSSM fields.
Even small couplings $g_{D,i}$ and $g_{L,i}$ of $\mathcal O(10^{-10})$ are sufficient 
in order for the messengers to decay rapidly.

The axino, as well as the saxion, are also thermalized at the temperature above
$10^9$\,GeV$(f_a/10^{11}{\rm GeV})^2$~\cite{Rajagopal:1990yx}.
The resulting abundance is similar to that of the gravitino.
The saxion constraint from the effective number of neutrino species is also marginal,
but it can easily be avoided by making the PQ scale slightly smaller, or by making the saxion heavy
as already described.

%%%%%%%%%%%%%%%%%%%%%%%%%%%%%%%%
\section{Conclusion}  \label{sec:conc}
%%%%%%%%%%%%%%%%%%%%%%%%%%%%%%%%%

In this paper we have constructed a gauge-mediation model that 
naturally overcomes the $\mu/B\mu$-problem without introducing a large CP angle.
The model is based on the $Z_3$ invariant singlet extension of the MSSM, the so-called NMSSM.
The only extension is the addition of the extra matter having a charge under a hidden gauge symmetry, as well as that of the SM gauge group.
In this setup, a desired value of the singlet VEV is induced to give a $\mu$-term
and the well-known domain wall problem in the NMSSM is solved due to the presence of 
quantum anomaly through the hidden gauge symmetry.
This is also consistent with the PQ solution to the strong CP problem.
Moreover, the relative sign of the $\mu$-term and the gaugino mass can be preferable
for explaining the muon anomalous magnetic moment.

We have also investigated cosmology of the present model.
The hidden gauge symmetry promises the existence of newly long-lived particles.
In particular, the hidden glueball may have interesting cosmological implications
since it may have a lifetime long enough to be a source of late-time entropy production.
For the glueballs to decay before BBN, the dynamical scale of hidden gauge symmetry
cannot be much smaller than 1~GeV.
The lightest states in the extra matter are also stable, but they efficiently annihilate at the 
hidden QCD phase transition, and are not of cosmological importance.
In the case of late-time entropy production by hidden glueballs, 
the reheating temperature can be as high as $10^9$\,GeV and (non-)thermal leptogenesis
scenario works even after the dilution by the entropy production is taken into account,
while the gravitino with mass of $\mathcal O(0.1)$\,MeV accounts for the present dark matter.

The present scenario has various interesting properties at low energy;
weakly coupled light scalars and a light neutralino originating from the singlet field, 
hidden gauge interaction with a dynamical scale around 10 GeV, 
and extra vector-like matter at TeV scale
charged under both standard model and hidden gauge symmetries.
These distinct features may be tested by the LHC or other experiments in the near future.

%%%%%%%%%%%%%%%%%%%%%%%%%%%%%%%%%%%%
\section*{Note Added}
\begin{figure}[tbp]
\begin{center}
 \includegraphics[scale=1.5]{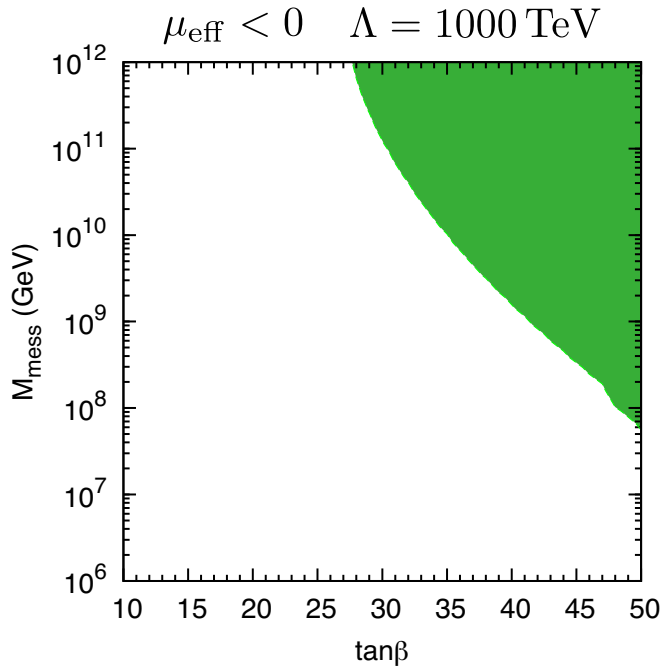}
\caption{The corresponding stop mass is $M_{\rm susy}\equiv \sqrt{m_{Q_3} m_{t_R}} \sim 10$ TeV, which slightly depends on the messenger scale. We take $\lambda(M_{\rm susy})=0.005$, $N_H=3$, $k_{L'}=0.002$, and $g_H(M_{\rm susy})=1.2$. The messenger number is taken as $N=1$.
}
\label{fig:highSUSY}
\end{center}
\end{figure}

%%%%%%%%%%%%%%%%%%%%%%%%%%%%%%%%%%%%
If we give up the explanation of the muon g-2 deviation, there  
 is a trivial solution which can explain the Higgs boson mass of $\sim125$ GeV with
 $\sim10$ TeV stops. In Fig.~\ref{fig:highSUSY}, the allowed region of the parameter space is shown for $ \mu_{\rm eff}<0$, and $\Lambda_{\rm eff}=1000$ TeV. Only the green region is excluded due to the constraint from the vacuum stability. In the other region, the Higgs mass of $\sim 125$ GeV is explained with $\sim$ 10 TeV stops. There exists an almost singlet-like (CP-even) scalar around $300-350$ GeV, which may be observed at the LHC. The solution with $\mu_{\rm eff}>0$ also exists. But the allowed region is rather small.

%%%%%%%%%%%%%%%

%%%%%%%%%%%%%%%%%%%%%%%%%%%%%%%%%%%%
\section*{Acknowledgment}
%%%%%%%%%%%%%%%%%%%%%%%%%%%%%%%%%%%%
NY would like to thank CERN TH group, 
and 
KH would like to thank DESY theory group and TUM HEP theory group, 
where part of this work has been carried out.
KN would like to thank S.~Shirai and M.~Kusakabe for useful conversations.
KH would like to thank M.~Ratz and M.~W.~Winkler for discussions.
This work is supported by Grant-in-Aid for Scientific research from
the Ministry of Education, Science, Sports, and Culture (MEXT), Japan,
No.\ 21111006 (K.N.), No.\ 22244030 (K.N.), No.\ 21740164 (K.H.), No.\ 22244021 (K.H.), and No.\ 22-7585 (N.Y.)
and by World Premier International Research Center Initiative (WPI Initiative), MEXT, Japan.

%%%%%%%%%%%%%%%%%%%%%%%%%%%%%%%%%%%%
\appendix
%%%%%%%%%%%%%%%%%%%%%%%%%%%%%%%%%%%%

\section*{Appendix A}
Two-loop beta functions for gauge couplings and gauginos are shown.
\begin{eqnarray}
(16\pi^2)\frac{d g_i}{dt} &=& b^{(1)}_i g_i^3 + \frac{g_i^3}{16\pi^2}
\left(\sum_{j=1}^4 b^{(2)}_{ij} g_j^2 - \sum_{x=t,b,\tau} c_{ix} Y_x^2
\right) , \nn
(8\pi^2)\frac{d M_i}{dt} &=& b^{(1)}_i g_i^2 M_i^2 +
\frac{g_i^3}{16\pi^2} \left(\sum_{j=1}^4 b^{(2)}_{ij} g_j^2(M_i+M_j)
-\sum_{x=t,b,\tau} c_{ix} Y_x^2(M_i-A_x) \right),
\end{eqnarray}
where $t=\ln Q$, $g_4=g_H$, and
\begin{eqnarray}
b^{(1)}_i = \left(
\begin{array}{c}
33/5 + N_H \\
1 + N_H \\
-3+ N_H \\
-3N_H + 5
\end{array}\right),
\end{eqnarray}
\begin{eqnarray}
b^{(2)}_{ij} =
\left(
\begin{array}{cccc}
199/25 + 23N_H/15 & 27/5 + 9N_H/5 & 88/5 + 32N_H/15 & 2(N_H^2-1) \\
9/5 + 3N_H/5 & 25+7N_H  & 24   & 2(N_H^2-1) \\
11/5 + 4N_H/15 & 9  & 14+34N_H/3  & 2(N_H^2-1) \\
2 & 6  & 16  & -6N_H^2+20N_H -10/N_H \nonumber
\end{array}
\right) ,
\end{eqnarray}
\begin{eqnarray}
c_{ix} =
\left(
\begin{array}{ccc}
26/5 & 14/5 & 18/5   \\
6 & 6& 2   \\
4 & 4 & 0   \\
0 & 0 & 0
\end{array}
\right).
\end{eqnarray}
Note that in the case of $N_H=3$, $b^{(1)}_3$ vanishes accidentally,
therefore the terms proportional to $b^{(2)}_{3j}$ and $c_{3x}$ are
important. We omit contributions proportional to Yukawa couplings
other than $Y_t$, $Y_b$ and $Y_\tau$, since they are small.

\providecommand{\href}[2]{#2}\begingroup\raggedright\endgroup

\end{document}